\documentclass[fleqn,usenatbib]{mnras}

\usepackage{newtxtext,newtxmath}

\usepackage[T1]{fontenc}

\DeclareRobustCommand{\VAN}[3]{#2}
\let\VANthebibliography\thebibliography
\def\thebibliography{\DeclareRobustCommand{\VAN}[3]{##3}\VANthebibliography}

\usepackage{graphicx}	
\usepackage{amsmath}	

\usepackage{enumitem}
\usepackage{xcolor}
\definecolor{ochre}{rgb}{0.8, 0.47, 0.13}

\usepackage[normalem]{ulem}

\title[Migration and mergers in massive star systems]{Formation of massive multiple-star systems: early migration and mergers}

\author[S. Chon \& Vigna-G\'omez]{
Sunmyon Chon$^{1}$\thanks{E-mail: sunmyon@mpa-garching.mpg.de} \&
Alejandro Vigna-G\'omez$^{2,1}$\thanks{E-mail: avignagomez@nbi.ku.dk}\thanks{Both authors contributed equally.}
\\
$^{1}$Max-Planck-Institut f\"ur Astrophysik, Karl-Schwarzschild-Str.~1, 85748 Garching, Germany\\
$^{2}$Niels Bohr International Academy, Niels Bohr Institute, University of Copenhagen, Blegdamsvej 17, 2100, Copenhagen, Denmark
}

\begin{document}
\label{firstpage}
\pagerange{\pageref{firstpage}--\pageref{lastpage}}
\maketitle

\begin{abstract}
Massive stars are often found in multiple systems, yet how binary-star systems with very close separations ($\lesssim$ au) assemble remains unresolved. We investigate the formation and inward migration of massive-star binaries in Solar-metallicity environments using the star-cluster formation simulation of Chon et al. (2024), which forms a $1200\,M_\odot$ stellar cluster and resolves binaries down to 1 au separation. Our results indicate that stars more massive than $2\,M_{\odot}$ predominantly assemble in binary or triple configurations, in agreement with observations, with member stars forming nearly coevally. In most of these systems, the inner binary hardens by one to three orders of magnitude and reaches a steady-state within the first $0.1\,$Myr. Notably, all binaries whose final separations are below 10 au are hardened with the aid of circumbinary discs, highlighting disc-driven migration as a key to produce tight massive binaries. We further find that binaries form with random inclinations relative to the initial rotation axis of the cloud, and that mutual inclinations in triple systems follow an isotropic distribution, implying that stochastic interactions driven by turbulence and few-body dynamics are crucial during assembly and migration. Finally, stars with $M>2\,M_{\odot}$ often undergo repeated merger events during cluster evolution, yielding extreme mass ratios ($q<0.1$). Some of these products may evolve into compact-object binaries containing a black hole or neutron star, including X-ray binaries and systems detectable by Gaia.
\end{abstract}



\section{Introduction}
Massive stars—typically those with masses $\gtrsim 8\,M_{\odot}$—play a fundamental role in shaping the evolution of galaxies. They drive chemical enrichment through nucleosynthetic yields from stellar winds and supernovae \citep[e.g.][]{2012ceg..book.....M, 2013ARA&A..51..457N, 2019A&ARv..27....3M, 2020MNRAS.496...80V}, inject radiative and mechanical feedback into their environments \citep[e.g.][]{2012MNRAS.421.3522H, 2013ApJ...770...25A, 2014prpl.conf..243K}, and ultimately end their lives in energetic explosions such as supernovae and gamma-ray bursts \citep[e.g.][]{2003ApJ...591..288H}. A large fraction of massive stars form and evolve in binary or higher-order multiple systems \citep{2012Sci...337..444S, 2013ARA&A..51..269D, 2017ApJS..230...15M, 2023ASPC..534..275O}. Some massive binaries later evolve into compact-object systems associated with high-energy transients and gravitational-wave (GW) emission. In particular, binary neutron-star mergers power short gamma-ray bursts and kilonovae \citep{2013ApJ...778L..16H, 2017ApJ...848L..12A}, and provide one major origin of $r$-process nucleosynthesis \citep{2017Natur.551...80K}, although additional channels—including collapsars—may also contribute significantly \citep{2019Natur.569..241S}. While binary black hole mergers do not play a major role in chemical enrichment, they account for the majority of GW detections to date \citep[e.g.,][and references therein]{2019PhRvX...9c1040A, 2021PhRvX..11b1053A, 2023PhRvX..13d1039A, 2024PhRvD.109b2001A}.

Despite their astrophysical importance, the formation of massive stars and the origin of their high multiplicity remain open questions \citep[e.g.,][]{2002ApJ...569..846Y,2007ARA&A..45..481Z, 2014prpl.conf..149T}. Massive star formation takes place within dense, optically thick regions that are accessible only at infrared or (sub-)millimetre wavelengths. The early phase is short-lived, characterised by rapid accretion, strong radiative feedback, and powerful outflows \citep[e.g.][]{2007prpl.conf..165B}, while the formation timescale for the most massive stars ($\sim$1–10 Myr) can constitute a significant fraction of their total lifetimes \citep{2014prpl.conf..243K}. These challenges limit our ability to determine how massive stars assemble and, particularly, how binaries form and dynamically evolve. Although multiplicity fractions among massive stars are high \citep[e.g.][]{2012Sci...337..444S, 2017ApJS..230...15M}, the physical processes that produce close binaries remain uncertain. Several mechanisms — including disc fragmentation, large-scale core or filament fragmentation, migration within massive discs, and dynamical interactions in young stellar environments — may all contribute, but their relative roles are poorly constrained. Understanding how massive binaries shrink from their initial separations to the very tight configurations observed today ($\lesssim$ few au) is therefore a major unresolved problem in massive star formation theory.

Recent high-resolution observations have begun to probe multiplicity in deeply embedded, high-mass protoclusters \citep[e.g.][]{2016Natur.538..483T}. In particular, \citet{2024NatAs...8..472L} identified binary and higher-order systems in a massive protocluster expected to form OB stars. These observations complement numerical simulations showing that massive stars predominantly form in clustered environments via the collapse of massive molecular clouds \citep[][and references therein]{2025ARA&A..63....1B}. However, the dynamical evolution of such systems—including the processes responsible for hardening them to tight separations—remains unclear. Observational studies suggest that massive binaries may undergo significant hardening within the first few Myr: \citet{2021A&A...645L..10R,2024A&A...690A.178R} reported a correlation between radial-velocity dispersion and cluster age, interpreted as evidence for early-time binary shrinking driven by interactions with residual discs or nearby stellar companions. Yet the relative importance of disc-driven torques, turbulent gas dynamics, and few-body interactions remains uncertain.

Observations further indicate that the distribution of massive binary separations is bimodal, with one peak at $\sim 10^{4}$ au and another at $\sim 1$–$100$ au \citep{2017ApJS..230...15M, 2023ASPC..534..275O}. This has motivated the interpretation that the two populations originate from distinct physical processes: wide binaries from large-scale core or filament fragmentation, and tighter systems from disc fragmentation. Theoretical models support this view, showing that massive circumstellar discs can become gravitationally unstable and fragment \citep[e.g.][]{2010ApJ...708.1585K}. Such fragments may subsequently migrate inward through disc torques, producing binaries with separations below $\sim 100$ au \citep[e.g.][]{2012ApJ...746..110Z, 2015ApJ...805..115V, 2018MNRAS.473.3615M, 2020A&A...644A..41O}. However, disc fragmentation alone may not explain the origin of very tight binaries ($\lesssim 10$ au). Alternative pathways—including migration in gravitationally unstable discs, multi-body interactions, and hierarchical triple evolution—may also play an important role, and binaries formed via large-scale fragmentation can undergo substantial orbital evolution \citep{2023A&A...674A.196K}. The formation of tight massive binaries therefore likely involves a combination of processes operating across a wide range of spatial scales.

Understanding these pathways requires simulations that combine large dynamic range with high spatial resolution, capable of modelling both cluster-scale collapse and binary-scale interactions. Resolving separations of order $\sim$1 au is essential for capturing binary hardening during the deeply embedded phases of massive star formation. Several recent studies have begun to explore this regime. \citet{2023MNRAS.518.4693G} analysed the binary populations formed in the STARFORGE suite \citep{2021MNRAS.502.3646G}, which follows the evolution of star-forming clouds over several Myr while resolving binaries down to $\sim 10$ au. \citet{2012MNRAS.419.3115B} performed long-term, high-resolution simulations of lower-mass clusters. These studies have provided valuable insights into the statistical properties and environmental dependence of young binaries. A complementary approach is to investigate the detailed dynamical pathways of individual massive binaries—particularly those that evolve toward the tightest separations.

In this work, we employ the high-resolution, Solar-metallicity star-cluster formation simulation presented in \citet{2024MNRAS.530.2453C} to investigate the formation and early evolution of massive binaries. The simulation follows the collapse of an initially $6300~M_{\odot}$ molecular cloud, producing a $\sim 1200~M_{\odot}$ stellar cluster while self-consistently incorporating radiative feedback from newly formed stars. Roughly one hundred binary systems form over the course of the simulation, spanning separations from $\sim 1$ to $10^{5}$ au. Using this dataset, we follow the assembly of binary and triple systems, quantify the hardening of gravitationally bound pairs, and analyse their orbital architectures—including inclinations and hierarchical structure. We also identify systems that may host stellar merger remnants or evolve into tight massive binaries. Our results provide new insights into the physical pathways through which massive stars migrate to the close separations required for the formation of compact-object binaries.

This paper is organised as follows.
In Section~\ref{sec:merger_correction}, we describe the numerical methodology, including the underlying star-cluster formation simulation and the post-processing procedures used to identify and characterise binary systems.
Section~\ref{sec:results} presents our main results on the formation, orbital evolution, and statistics of binaries found in the simulation.
In Section~\ref{sec:discussion} we discuss the physical interpretation and broader implications of these findings.
We summarise our conclusions in Section~\ref{sec:conclusion}.

\section{Methodology} \label{sec:methodology}
Our study utilises a high-resolution simulation of a star-forming cloud to investigate the formation processes of tight binary systems.
A full description of the base simulation is available in \cite{2024MNRAS.530.2453C}.
This section is organized as follows: in Section~\ref{sec:simulation}, we briefly describe the key assumptions of the base simulation, and in Section~\ref{sec:postprocessing}, we outline the methodology used for the analysis presented in this paper.

\subsection{Base Simulation} \label{sec:simulation}
The star formation process is followed by using the smooth-particle hydrodynamic (SPH) simulation code, {\tt Gadget3} \citep{2005MNRAS.364.1105S}, with extensions presented in \citet{2021MNRAS.508.4175C} and \citet{2024MNRAS.530.2453C} that capture the physics relevant to the star formation, such as the detailed chemistry and thermal evolution, particle splitting, sink particle formation and accretion, and the radiation transfer of the ultra-violed (UV) and far-UV photons using ray-tracing method. 
We will focus exclusively on the simulation at Solar metallicity.
Here we summarise the fundamental physics and key assumptions underlying the base simulation.

\begin{description}
    \item \textbf{Initial Conditions.}
    The simulation begins from a rotating and turbulent Bonnor-Ebert sphere \citep{1956MNRAS.116..351B} with the density $n = 10^4~\mathrm{cm^{-3}}$ and temperature $T = 200~$K. 
    The initial mass and the size of the cloud is $6300~M_\odot$ and $1.2\times 10^6~\mathrm{au} = 5.9~$pc, respectively. 
    We assume a rigid rotation with $\Omega = 2.08\times10^{-15}~ \mathrm{s^{-1}}$, where the rotational energy is $0.1\%$ of the gravitational energy, consistent with the observation of Galactic molecular cloud cores \citep{2002ApJ...572..238C}. To model turbulent motions, we impose a transonic turbulence with a velocity dispersion $v_\text{disp} = c_\text{s}$, where $v_\text{disp}$ is the mass-weighted velocity dispersion of the gas, and $c_\text{s}$ is the sound speed \citep{1985MNRAS.214..379L}. The turbulent velocity field is initialised on a $128^3$ grid covering the simulation box surrounding the initial cloud. Gas particle velocities are interpolated from this grid using the cloud-in-cell (CIC) method. To construct the turbulent field, we seed random Fourier modes following a power spectrum consistent with Larson’s law, $P(k) \propto k^{-2}$ \citep{1999ApJ...524..169M}. An inverse Fourier transform is then applied to obtain the velocity field in real space.
    \item \textbf{Resolution.}
    Starting from the initially turbulent cloud core, \cite{2024MNRAS.530.2453C} studied the gravitational collapse, fragmentation, and the formation of the individual stars. The initial gas particle mass is $4.7\times 10^{-3}~M_\odot$.
    A two-level particle splitting method following \citet{2002MNRAS.330..129K} is adopted, where a particle is split into 13 daughter particles once the density exceeds $10^5$ and $10^8~\mathrm{cm^{-3}}$.
    This allows us to resolve down to the density of $n = 2\times 10^{15}~\mathrm{cm^{-3}}$ with splitting SPH particles to the mass of $5\times 10^{-5}~M_\odot$. 
    This results in resolution down to $1~$au in spatial and $\sim 0.01~M_\odot$ in mass scales \citep{1995MNRAS.277..362B}.
    \item \textbf{Sink Particles.}
    A sink particle is inserted once the gas density exceeds $2 \times 10^{15}~\mathrm{cm^{-3}}$. 
    Each sink particle represents an individual star.
    The sink particle may accrete gas particles around and grow in mass.
    Sink particles merge once the separation of a pair of them becomes smaller than the sum of the sink radii.
    To reduce the computational cost, we set sink radii increasing with stellar masses, 
    \begin{align}
    r_\text{sink} (M_*) = \max \left\{ 1.2 \sqrt{M_*/M_\odot},\ 0.85 \right \}~\mathrm{au}.
    \end{align}
    This sets the spacial resolution of our simulation, especially the minimum separation of binaries.
    Note that the gravitational softening for sink particles is set to be $0.2~$au, which is much smaller than the minimum sink radius.
    \item \textbf{Cooling.}
    Non-equilibrium evolution of the thermal energy is followed. 
    Several cooling processes are implemented, e.g., line-cooling by atomic hydrogen, molecular hydrogen, HD, CII, and OI, dust thermal emission, and continuum process \citep{2021MNRAS.508.4175C}.
    We estimate the cooling rate by dust thermal emission, solving the detail balance of the cooling/heating of dust grains, including dust-gas collision, heating by cosmic microwave background radiation and local stellar irradiation \citep{1979ApJS...41..555H}.
    \item \textbf{Feedback.}
    Radiation feedback is implemented from massive stars with masses exceeding $10~M_\odot$. 
    The radiative processes considered include hydrogen photoionization and the associated photoheating, molecular hydrogen dissociation, photoelectric heating, and dust heating. 
    The rates of photoionization, molecular dissociation, and photoelectric heating are computed using the ray-tracing scheme RSPH \citep{2006PASJ...58..445S,2017MNRAS.467.4293C}. 
    Dust heating is modelled by assuming that the radiative flux from a star decreases as $1/r^2$ with distance, which well-reproduces the dust temperature distribution obtained in multi-wavelength radiation hydrodynamic simulations \citep{1999ApJ...525..330Y, 2018A&A...616A.101K, 2020MNRAS.497..829F}. 
    Note that we do not include other feedback processes like jet or wind feedback, nor SNe feedback. We will further discuss these caveats and possible impact of those effects in Section~\ref{sec:discussion_multiplicity}.
\end{description}

\subsection{Post-Processing} \label{sec:postprocessing}
We have analysed 11,000 snapshots to extract binary properties, with outputs recorded every 1,000 computational timesteps. The typical time interval between snapshots is approximately $180~\mathrm{yr}$, although it varies slightly depending on the local timestep criteria. 
In this section, we describe our approach to identifying binary and higher-order multiple systems, determining the hardening mechanism, and correcting for merger effects given our spatial resolution.

\begin{figure}
    \centering
    \includegraphics[width=\hsize]{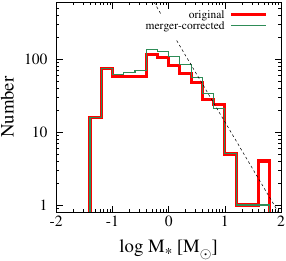}
    \caption{
    Mass spectrum obtained from the base simulation. 
    The red thick line shows the spectrum from the original data and the green thin line shows the distribution of the ``merger-corrected'' sample described in Section~\ref{sec:merger_correction}.
    The dashed line shows the Salpeter IMF with $\mathrm{d}N/\mathrm{d}M_* \propto M_*^{-2.3}$ \citep{1955ApJ...121..161S}.
    }
    \label{fig:IMF}
\end{figure}

\subsubsection{Determination of stellar multiplicity}
We identify binary and higher-order stellar systems using the following procedure. 
At the end of the base simulation, a total of 750 sink particles are present. 
Binary identification follows the method of \citet{2012MNRAS.419.3115B}. 
We begin by identifying the most gravitationally bound pair of sink particles. 
This pair is then replaced by a single particle whose mass is the sum of the two components, and whose position and velocity correspond to the centre of mass of the pair. 
This procedure is iteratively applied until no further gravitationally bound pairs are found. 
To limit the complexity of multiple systems, we impose a maximum multiplicity of three. 
That is, once a bound group reaches three members, it is removed from further consideration in the sink particle list.

\subsubsection{Classification of binary formation modes}
We classify binary formation into four distinct modes based on the morphology of the parent gas clouds at the time of the birth of the binary: filament fragmentation, disc fragmentation, core fragmentation, and dynamical capture \citep[e.g.][]{2023ASPC..534..275O}.
The formation time of a binary system is defined as the moment when the sink particles representing the binary components are created.

To analyze the gas structures associated with binary formation, we identify the relevant region at the time of binary birth using the following procedure.
We begin by defining a threshold density $n_\text{th}$ as the highest density at which the two binary components are still part of a connected gas structure.
To determine this, we perform a friends-of-friends (FoF) search on gas particles with densities exceeding a variable threshold $n_\text{FoF}$, using the smoothing length of each particle as the local linking length.
Starting from the gas particles nearest to each of the two binary components, we test whether they are part of the same FoF group at a given density.
We define $n_\text{th}$ as the maximum density at which this connection exists.
Using this threshold, we then extract the gas particles with 
the density larger than $n_\text{th}$ using the FoF algorithm, and define the resulting structure as the ``binary progenitor cloud''.
To characterize its morphology, we approximate the cloud as a non-axisymmetric ellipsoid by calculating the second moment of the mass distribution:
\begin{align}
I_{i,j} = \sum_p (x_{p,i}-\bar{x}_{p,i})(x_{p,j}-\bar{x}_{p,j}),
\end{align}
where the sum runs over all the particles $p$ in the binary progenitor cloud, and $i, j \in \{x, y,z \}$. 
We then diagonalize the moment-of-inertia tensor $I_{i,j}$ to obtain its eigenvalues, which correspond to the squared lengths of the three principal axes of the ellipsoid.
Finally, we define $\lambda_1$, $\lambda_2$, and $\lambda_3$ as the square roots of the eigenvalues, ordered such that $\lambda_1 \ge \lambda_2 \ge \lambda_3$.

We determine the dynamical capture mode occurs when the threshold density $n_\text{th}$ is lower than the mean density of the initial cloud, which is $10^4~\mathrm{cm^{-3}}$.
This criterion ensures that the formation sites of the binary components are completely disconnected from each other and from the natal cloud.
When $n_\text{th}$ exceeds the mean density of the initial cloud, we classify the formation modes based on the axis ratios derived from $\lambda_1$, $\lambda_2$, and $\lambda_3$. 
If $\lambda_1 > 4 \lambda_2$, we define the progenitor cloud as having a filament-like morphology and classify the formation mode as filament fragmentation. 
If $\lambda_1 < 4 \lambda_2$ and $\lambda_2 > 4 \lambda_3$, we interpret the progenitor cloud as disc-like and classify the mode as disc fragmentation. 
All other cases are classified as core fragmentation.
This classification agrees well with an independent visual (human-eye-based) inspection.

\subsubsection{Determination of "hardening" mechanism} \label{sec:merger_correction}
To evaluate the importance of interactions with the circumstellar disc, we determine whether the binary members are embedded in or interacting with their respective discs. 
We first estimate the disc radius by comparing the rotational velocity ($v_\text{rot}$) to the Keplerian velocity ($v_\text{Kep}$), defined as
\begin{align}
v_\text{Kep}(r) \equiv \sqrt{\frac{G M(<r)}{r}},
\end{align}
where $G$ is the gravitational constant, $r$ is the distance from a given star, and $M(<r)$ is the enclosed mass within radius $r$ from the star. 
We define the disc radius as the radius beyond which the ratio 
$v_\text{rot} / v_\text{Kep}$ drops below $0.7$.
If the distance between the two component stars of the binary becomes smaller than the sum of their individual disc radii, we consider the system to be in a phase of disc-star interaction, hereafter referred to as disc interaction.
Once the disc radius becomes larger than twice the binary separation, we define the disc as a circumbinary disc.

\begin{figure*}
    \centering
    \includegraphics[width=\hsize]{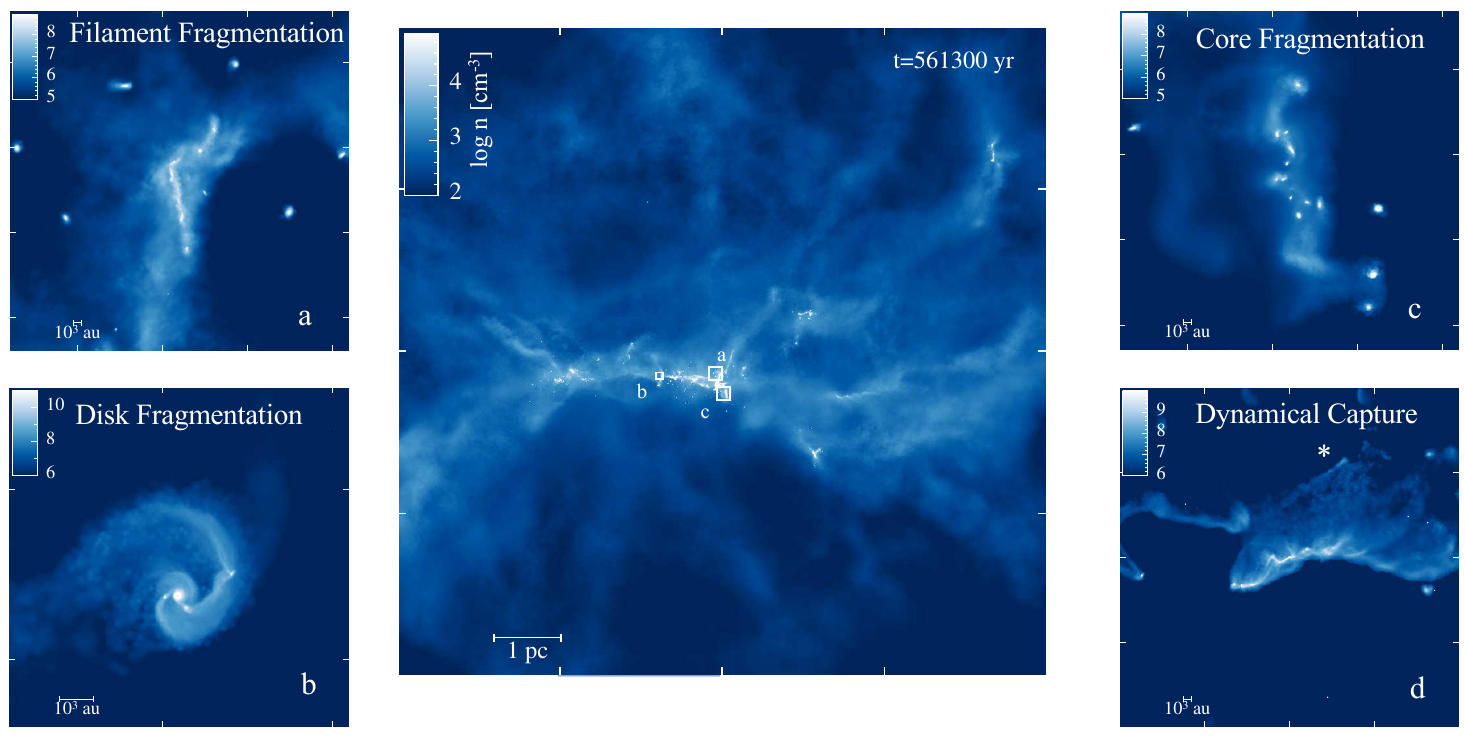}
    \caption{
    Snapshot of the base simulation, showcasing examples of the key mechanisms driving multiple formation.
    Central panel: projected density distribution for the entire cloud scale at $t\approx0.56~$Myr.
    Panels a-d: zoom-in view of the binary progenitor clouds for four different binary formation channels, filament fragmentation (a), disc fragmentation (b), core fragmentation (c), and dynamical capture (d). The white squares a-b in the central panel indicate the regions where binary formation occurs.
    We do not specify the region of the binary formation by dynamical capture (panel d) because it occurs at $t\approx1.7~$Myr, far later than the shown snapshot.
    A scale bar representing $1$ pc (or $10^3$ au) is shown at the bottom-left part of the central panel (or panels a-d).
    Note that the colour-scale is different in each panel.
    }
    \label{fig:overall}
\end{figure*}

\subsubsection{Correction for mergers} \label{subsubsec:correct_mergers}
One major uncertainty in estimating the multiplicity fraction arises from how stellar mergers are numerically handled. In the base simulation, two stars merge when their separation falls below the sum of their sink radii, which are typically in the range of $\sim 1$--$10~\mathrm{au}$. However, the actual stellar radii are on the order of $1$--$10~R_\odot$. This means some systems might survive as tight binaries instead of physically merging. To assess this uncertainty, we construct an alternative "merger-corrected" stellar sample in which all merged stars are assumed to have survived as tight binaries.

During the simulation, 118 merger events are recorded. We reconstruct a corrected sample by treating these merged systems as distinct, unresolved binaries. When added to the original 739 sink particles, the merger-corrected sample comprises 857 stars. The stellar masses are adjusted by subtracting the mass contributions from the merger events to better reflect their pre-merger states. Note that, by neglecting any post-merger mass accretion, this correction may overestimate the primary\footnote{Throughout this manuscript, "primary" denotes the more massive star in a binary system or, in a triple system, the more massive star within the inner binary, unless otherwise noted.} stellar masses.

Figure~\ref{fig:IMF} presents the mass spectrum of our original sample (thick red line) with that of the merger-corrected sample (thin green line). The correction has a minor impact on the overall shape of the mass function, aside from the high-mass end. The dashed line represents the Salpeter initial mass function (IMF) with a slope of $-2.3$, and both distributions are consistent with it at $M_* \gtrsim 2~M_\odot$.

\begin{figure*}
    \centering
    \includegraphics[width=\hsize]{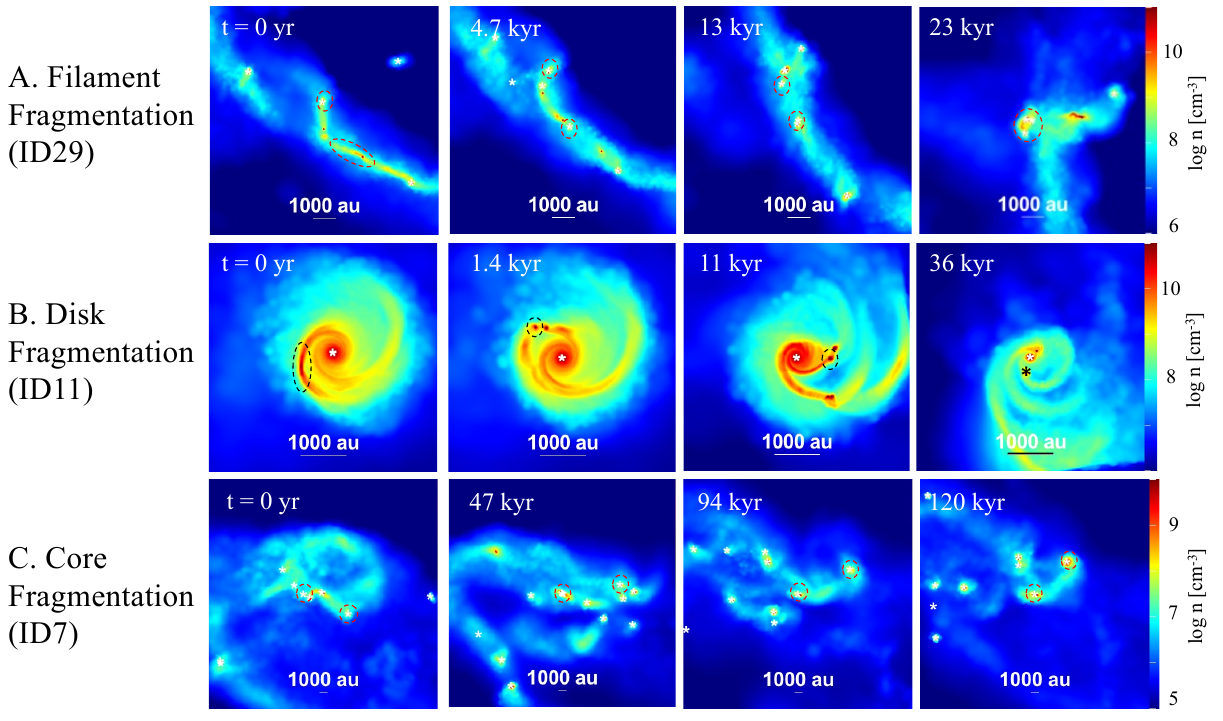}
    \caption{
    Examples of the binary formation and evolution for different binary formation channels, filament fragmentation (A), disc fragmentation (B), and core fragmentation (C).
    We show the projected density distributions of the binary progenitor clouds.
    The time origin indicates when the fragmentation occurs.
    We denotes the member of the binary system or its progenitor cores by dashed circles.
    }
    \label{fig:binary_fig_individual}
\end{figure*}

\begin{table*}
\caption{
Summary of the statistics of key mechanisms driving multiple formation for all binary star systems from our base simulation. The percentage is calculated relative to the total number of systems within each (sub)sample.
}
\label{tab:formation_mode}
\begin{tabular}{c|c|cccc}\hline
 & All & Filament & Disc & Core  & Dynamical \\
  & Modes & Fragmentation & Fragmentation & Fragmentation  & Capture \\ \hline
All Binaries & 76 & 29 (38\%) & 16 (21\%) & 20 (26\%) & 11 (14\%) \\
$a_\text{final} < 100~$au & 49 & 24 (49\%) & 11 (22\%) & 14 (29\%) & 0 (0\%) \\ 
$a_\text{final} \ge 100~$au & 27 & 5 (18\%) & 5 (18\%) & 6 (22\%) & 11 (41\%) \\ 
$a_\text{final} \ge 10^4~$au & 11 & 0 (0\%) & 0 (0\%) & 1 (9\%) & 10 (91\%) \\ 
\hline
\end{tabular}
\end{table*}

\section{Results} \label{sec:results}
\subsection{Overview}
Figure~\ref{fig:overall} shows the projected density distribution of the simulated cloud at time $t\approx 0.5~$Myr after the formation of the first protostar\footnote{Throughout this manuscript, the formation of this first protostar determines the reference point we use to report timescales, unless otherwise noted.}. The chosen snapshot is the typical formation epoch of the stars, when the star formation rate peaks \citep{2024MNRAS.530.2453C}.
On spatial scales of $\lesssim 10~$pc, a long, turbulent filament with the density of $\sim 10^{5}~ \mathrm{cm^{-3}}$ dominates the cloud. Self-gravity and turbulence carve this structure into sub-structures, such as, cores, discs, and secondary filaments, which then collapse into protostars that later form binaries or higher-order multiples.

Following \citet{2023ASPC..534..275O}, Fig.~\ref{fig:overall} illustrates the four primary binary-birth channels in the following panels: A) filament fragmentation, in which a dense filament with the density of $10^8~\mathrm{cm^{-3}}$ fragments into protostars along its axis; B) disc fragmentation, in which a massive circumstellar disc, shown face-on, becomes gravitationally unstable and spawns companions near its spiral arms; C) core fragmentation, in which a compact core with the density of $10^{6}~\mathrm{cm^{-3}}$ splits into a handful of protostars, often forming a mini-cluster; and
D) dynamical capture, in which stars that form in separate cores later pass within a few hundred au and become gravitationally bound. 
Together, and sometimes combined, these pathways account for the entire binary population in the simulation.

Panels A and C share the same line of sight as the large-scale density map, whereas panel B is viewed perpendicular to the large-scale filament. 
Panel D is taken from the later time at $t=1.7~$Myr, because no binary forms via capture before $t\sim 0.5~$Myr.
The fact that protostars in the filament- and core-fragmentation modes form predominantly perpendicular to the parent filament suggests that local turbulent motion, rather than the global filament geometry, governs small-scale collapse.

Figure~\ref{fig:binary_fig_individual} illustrates how the density distributions evolve over time for the various binary formation channels:
\begin{description}
    \item \textbf{Top row: filament fragmentation.} The initial high-density filament with the density of $10^8~\mathrm{cm^{-3}}$ fragments into a few spherical cores indicated by dashed circles. 
    These cores approaches each other along the direction of the filament and experience mergers. 
    After $23~$kyr from the fragmentation, the separation of the protostars reaches a few $100~$au, which marks the first pericentric passage.
    The timescale of the initial migration to hit the pericenter is a few $10~$kyr, comparable to the dynamical time of the initial filament
    \begin{equation}
    t_\text{dyn} = \frac{1}{\sqrt{G \rho}} = 8.8\times \left ( \frac{n}{10^8~\mathrm{cm^{-3}}} \right )^{-1/2} ~\mathrm{kyr},
    \label{eq:t_dynamical}
    \end{equation}
    indicating that the migration is caused by self-gravity of the filament.
    The evolution of the binary separation afterwards is driven by the interaction with the circumstellar disc and dynamical interactions with nearby cores and protostars.
    \item \textbf{Second row: disc fragmentation.} Spiral arms are waken at the circumstellar disc due to the self-gravity of the disc. One of the arm become further gravitationally unstable to fragments into a few spherical cores, which is shown at $t=1.4~$kyr. 
    These fragments migrate toward the central star via interaction with the disc, which is similar to ``Type-I migration'' \citep{1980ApJ...241..425G, 2002ApJ...565.1257T}, which often explains for early binary migration \citep{2012ApJ...746..110Z, 2019MNRAS.488.2658C}.
    Mergers consequently occur during the course of migration and ultimately a binary system forms.
    The angular momentum is removed by ejecting mass waking highly non-axisymmetric spiral arms, which are seen at $t=36~$kyr.
    \item \textbf{Third row: core fragmentation.} The initial ellipsoidal core with a density of $10^6~\mathrm{cm^{-3}}$ and a size of $\sim10^4~$au fragments into more than ten small cores or protostars. 
    Cores gravitationally interact with one another, eventually causing two protostars to become gravitationally bound and form a binary system.
    The timescale for forming a binary is $\gtrsim 120~$kyr and longer than the former cases due to a stochastic nature of the interaction. 
\end{description}

We do not include an illustrative example of dynamical capture resulting in binary formation. In this case, the protostars form in separate cores with an initial separation of approximately $\sim 10^4~$au. They then approach and become gravitationally bound about $10$--$100$ kyr after the formation of the initially fragmented cores.

\subsubsection{Statistics of binary formation modes}
Table~\ref{tab:formation_mode} summarises the statistics of binary formation modes for binaries\footnote{Throughout the manuscript, ``binary" refers to both isolated binaries and the inner components of triple star systems. When discussing the outer binary in a triple system, this is specified explicitly.} at all separations (top), following subsets of the sample filtered by final separations below and above $100~$au (second and third rows, respectively), and greater than $10^4~$au (the bottom row). 
For the general sample, the four formation channels contribute almost equally, with a slight excess in filament fragmentation and a mild deficit in dynamical capture.  
In contrast, for tight binaries with $a_\text{final} < 100~$au, the contribution from dynamical capture is zero and half the binary form via filament fragmentation.
In the case of wide binaries, the contribution of the dynamical capture increases, i.e. around $40\%$ of the binaries for $a_\text{final} \geq 100~$au and $90\%$ for $a_\text{final} \geq 10^4~$au. 

\begin{figure}
    \centering
    \includegraphics[width=\hsize,trim={1.5cm 2.75cm 2.5cm 1.0cm},clip]{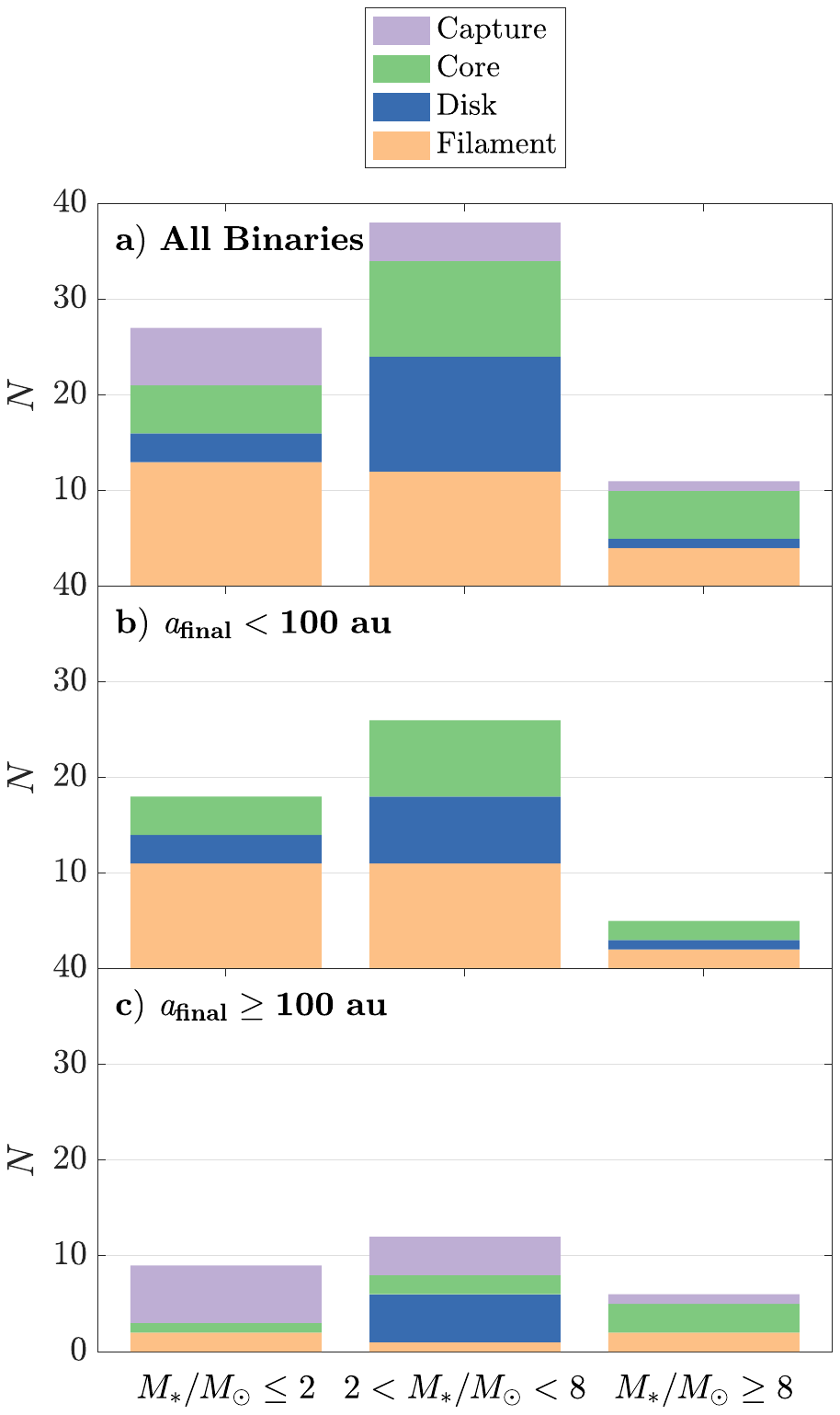}    
    \caption{
    The number of surviving isolated and inner binary systems as a function of primary stellar mass is shown for (a) all binaries, (b) binaries with final separations smaller than $100~\mathrm{au}$, and (c) binaries with final separations larger than $100~\mathrm{au}$. Different colour bins indicate distinct binary formation channels: orange for filament fragmentation, blue for disc fragmentation, green for core fragmentation, and purple for dynamical capture.
    }
    \label{fig:fragmentation_mass}
\end{figure}

Figure~\ref{fig:fragmentation_mass} shows histograms of the contributions from each binary formation channel as a function of primary stellar mass for (a) all binaries, (b) the subset of binaries with $a_\text{final} < 100~$au and (c) $a_\text{final} \geq 100~$au.  
We observe a clear trend: filament fragmentation is the dominant formation mechanism for binaries with low-mass primaries ($M_*/M_{\odot}\leq 2$), while disc fragmentation is predominant for binaries with intermediate-mass primaries ($2 < M_*/M_{\odot} < 8$).
For binaries with high-mass primaries ($M_*/M_{\odot}\geq 8$), core fragmentation becomes the dominant formation mode.  
Binary formation via dynamical capture contributes only a minor fraction across all mass ranges.

The underlying physical picture is as follows: as the stellar mass increases, the circumstellar disc also becomes more massive and increasingly susceptible to gravitational fragmentation, which accounts for the larger contribution of disc fragmentation for the intermediate mass stars than the low-mass stars. 
However, once the stellar mass exceeds $\sim 8~M_\odot$, the system tends to accumulate more gas under strong turbulent conditions.  
In this regime, ordered structures such as discs or filaments are largely absent, and the binary progenitor clouds can fragment in a more chaotic manner due to the combination of the self-gravity and turbulence, leading to core fragmentation.

For wide binaries with the final separation larger than $100~$au, the contribution of dynamical capture is $67\%$ for the low-mass stars, while it decreases for higher-mass systems. Only one binary system is formed via capture for $M_* > 8~M_\odot$. The reason for almost negligible contribution of capture is that those massive stars usually form in dense environments and numerous stars form out of the same progenitor cloud. This increases the probability of forming binary systems with nearby stars from the same progenitor cloud, as opposed to capturing stars formed in a different cloud.

\begin{figure*}
    \centering
    \includegraphics[width=\hsize]{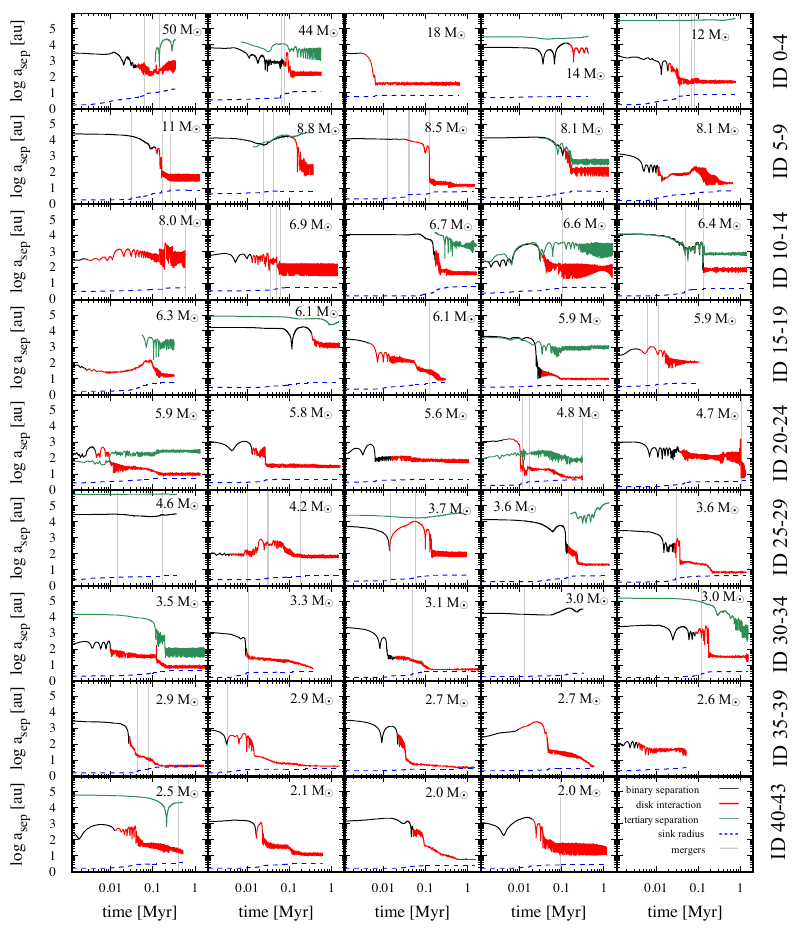}
    \caption{
    Time evolution of the orbital separation ($a_{\rm sep}$), measured from the binary's formation epoch.
    The time interval during which the binary interacts with a disc is highlighted in red. 
    The system’s spatial resolution---calculated as the sum of the sink radii of the binary---is indicated by a dashed blue line. 
    If the binary has a tertiary companion, its separation is shown as a solid green line.
    Merger events between sink particles are marked by thin solid grey vertical lines.
    Result for our base simulation at Solar metallicity with feedback included.
    }
    \label{fig:separation_evolution}
\end{figure*}

\begin{figure*}
    \centering
    \includegraphics[width=\hsize]{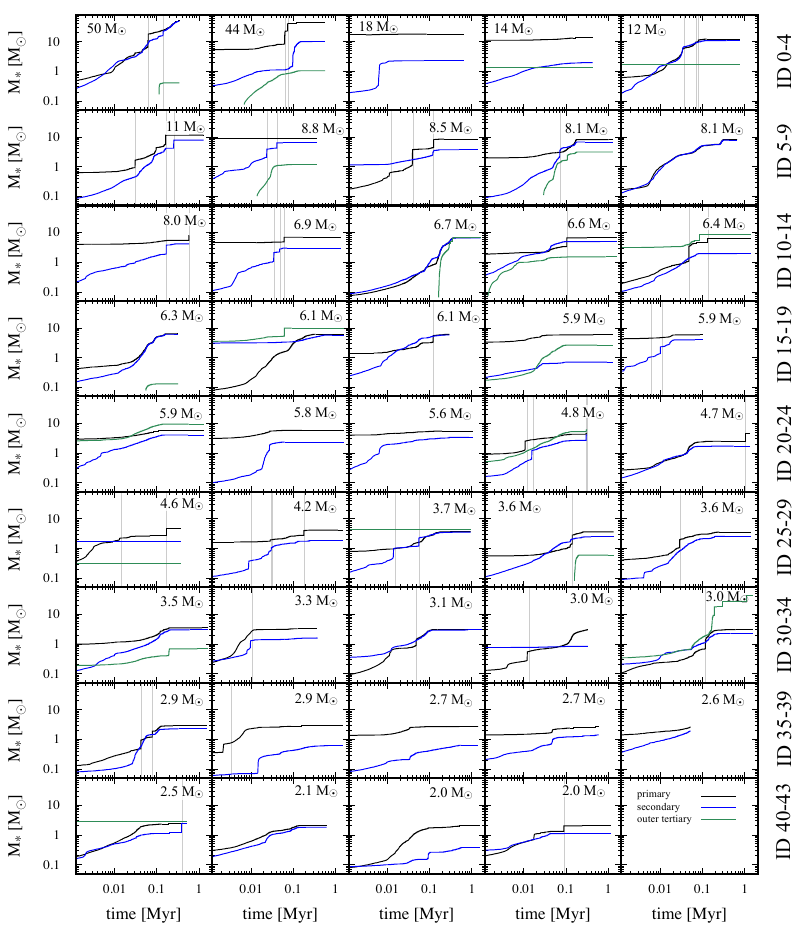}
    \caption{
    Time evolution of the mass of the star ($M_{*}$), measured from the binary formation epoch.
    We follow the mass accretion history of the primary (solid black), the secondary (solid blue), and, when present, the outer tertiary (solid green).
    Merger events between sink particles are marked by thin solid grey vertical lines.
    Result for our base simulation at Solar metallicity with feedback included.
    }
    \label{fig:mass_accretion_analysis}
\end{figure*}

\begin{figure*}
    \centering
    \includegraphics[width=\hsize]{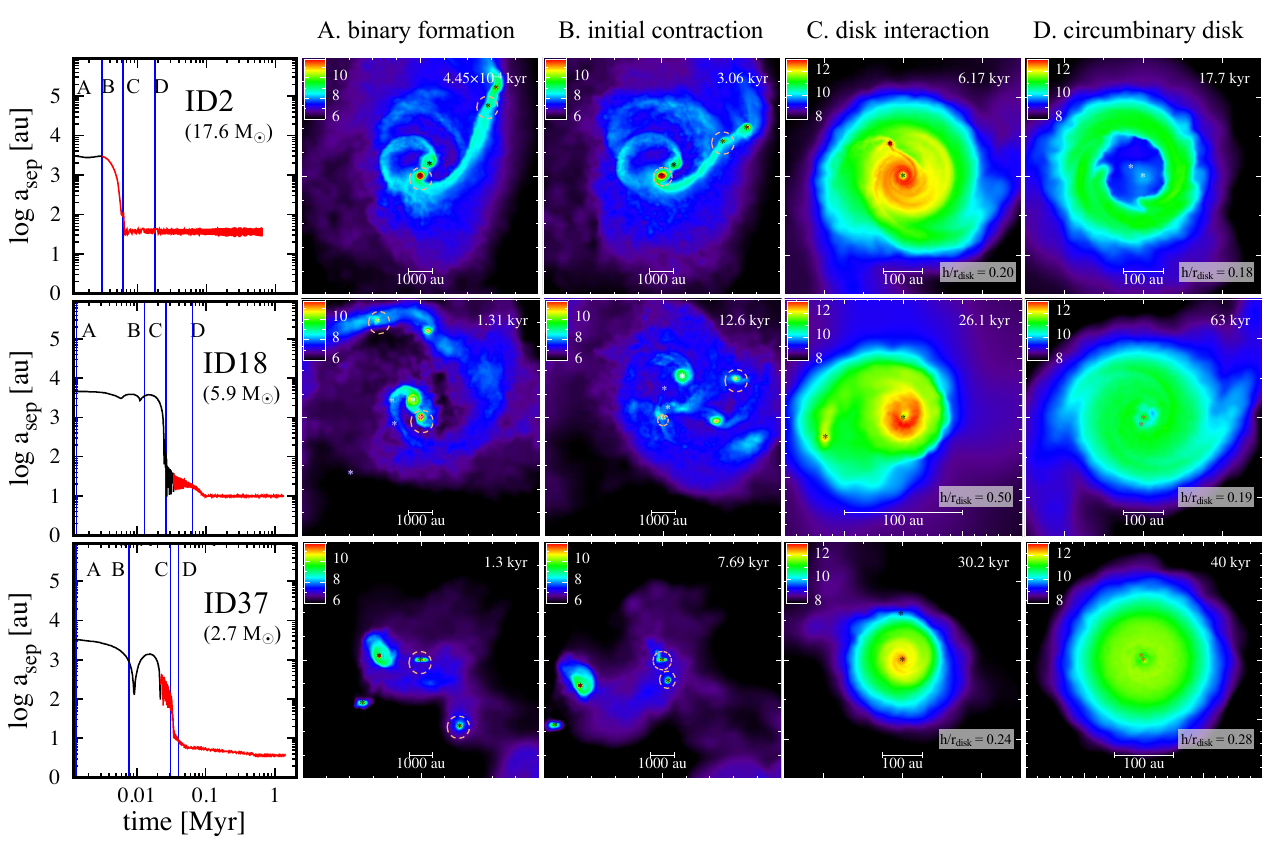}
    \caption{
    Example systems that form isolated binaries.
    Left column: time evolution of the binary separation as a function of time since binary assembly. The epoch corresponding to the disc interaction or circumbinary phases is highlighted in red. The timing corresponding to Panels A--D is marked by vertical blue solid lines.
    Panels A--D: projected gas density distributions around the binary member stars at four key evolutionary stages: binary formation (A), initial contraction phase (B), disc interaction phase (C), and circumbinary disc phase (D). Asterisks indicate the positions of protostars, and the binary members are highlighted by dashed circles in Panels A and B.
    At the end of the simulation, the eccentricity and separation of the binaries are 0.60 and $\approx 42$ au (ID2, top panel), 0.06 and $\approx9$ au (ID 18, middle panel), and 0.04 and $\approx 4$ au (ID 37, bottom panel), respectively.
    We also show the ratio of the scale height $h$ to the disc radius $r_{\mathrm{disc}}$ for stages (C) and (D).
    }
    \label{fig:circum_binary_fig}
\end{figure*}

\begin{figure*}
    \centering
    \includegraphics[width=\hsize]{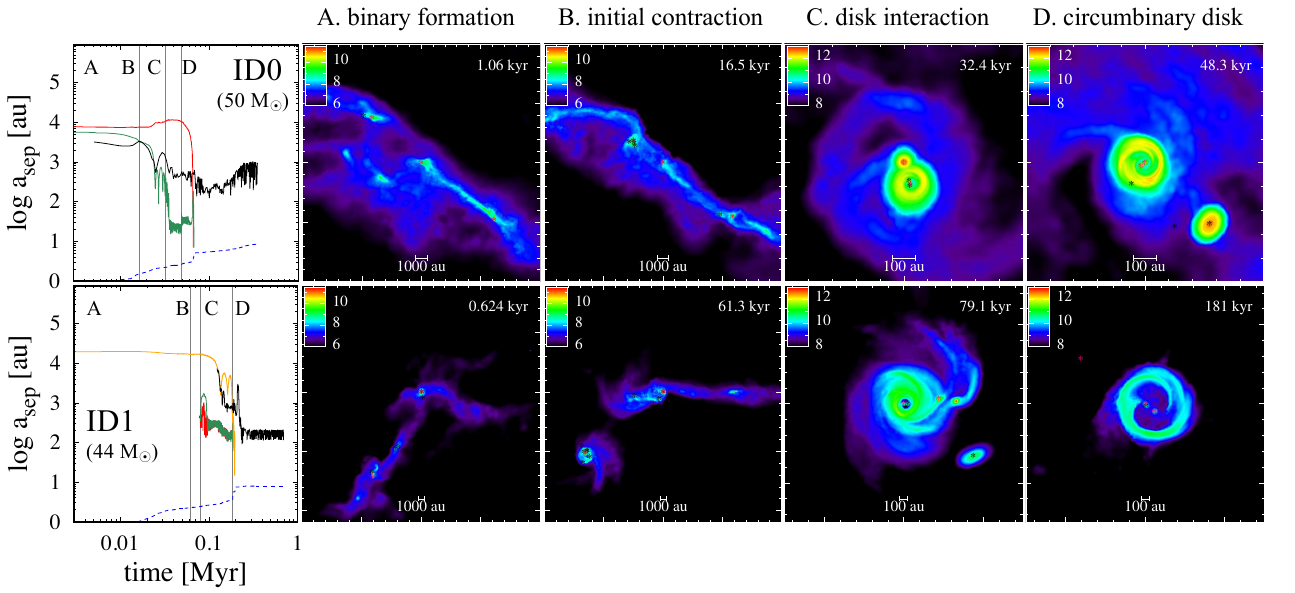}
    \caption{
    Example systems which undergo multi-body interactions and mergers. In the left column, the time evolution of the separations of the merged stars is shown by the red, green, and yellow lines. In panels A–D, stellar positions are overplotted as asterisks on the projected density distribution. Magenta asterisks denote stars that remain in the final three-body system or merged with the primary star, while black asterisks mark stars that are eventually ejected from the system.
    }
    \label{fig:merged_binary_fig}
\end{figure*}

\begin{figure}
    \centering
    \includegraphics[width=\hsize,trim={0.86cm 9.58cm 2.5cm 8.25cm},clip]{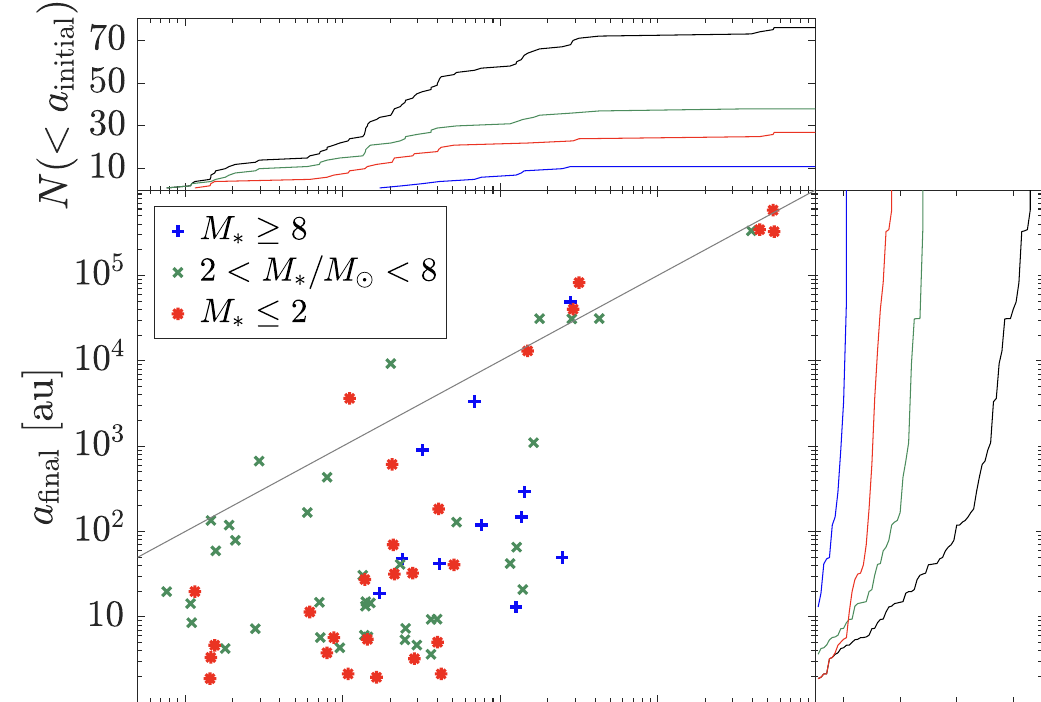}    
    \includegraphics[width=\hsize,trim={1cm 7.5cm 2.5cm 11.47cm},clip]{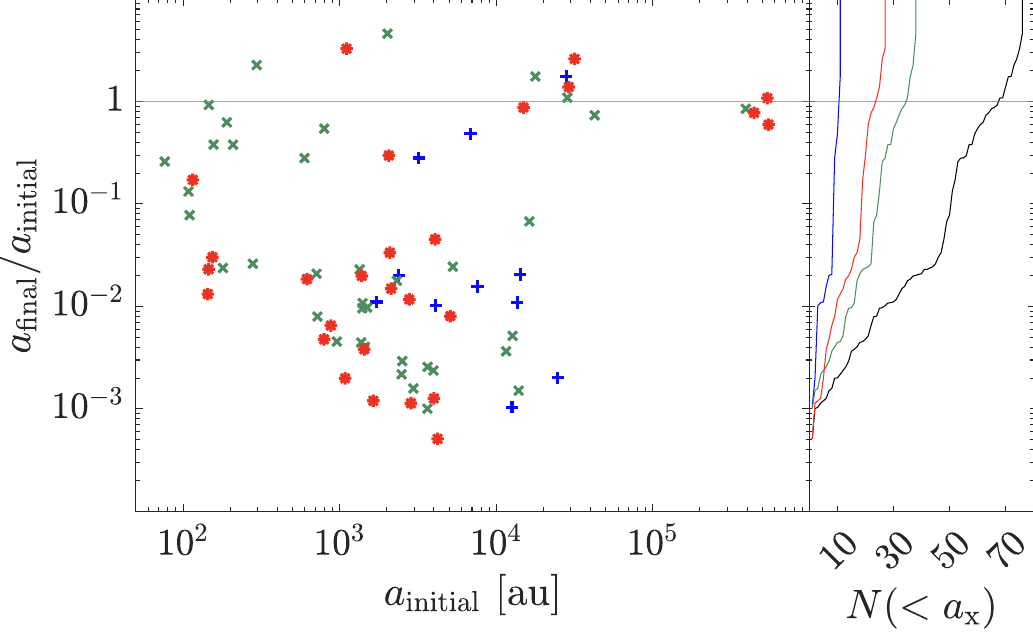}   
    \caption{
    Surviving isolated and inner binaries in our simulation.
    Top panel: Final separation ($a_{\rm final}$) versus initial separation ($a_{\rm initial}$).
    Bottom panel: Ratio $a_{\rm final}/a_{\rm initial}$, as a measure of binary hardening, versus $a_{\rm init}$.
    In each panel, the dashed grey line indicates $a_{\rm final}=a_{\rm initial}$ (diagonal in the top panel and horizontal in the bottom panel). 
    Different mass regimes are highlighted.
    Binaries in which the most massive star is $M_{*}/M_{\odot} \geq 8$ are shown as blue plus signs. 
    Binaries in which the most massive star is $2 < M_{*}/M_{\odot} < 8$ are shown as green crosses. 
    Binaries in which the most massive star is $M_{*}/M_{\odot}\leq 2$ are shown as red asterisks.
    The cumulative number of systems is shown along the top and right axes; the black line indicates all system, and the coloured lines follow the aforementioned mass regimes. 
    Result for our base simulation at Solar metallicity with feedback included.
    }
    \label{fig:ainit_afinal}
\end{figure}

\begin{figure}
    \centering
    \includegraphics[width=\hsize,trim={1cm 7.5cm 2.5cm 8.25cm},clip]{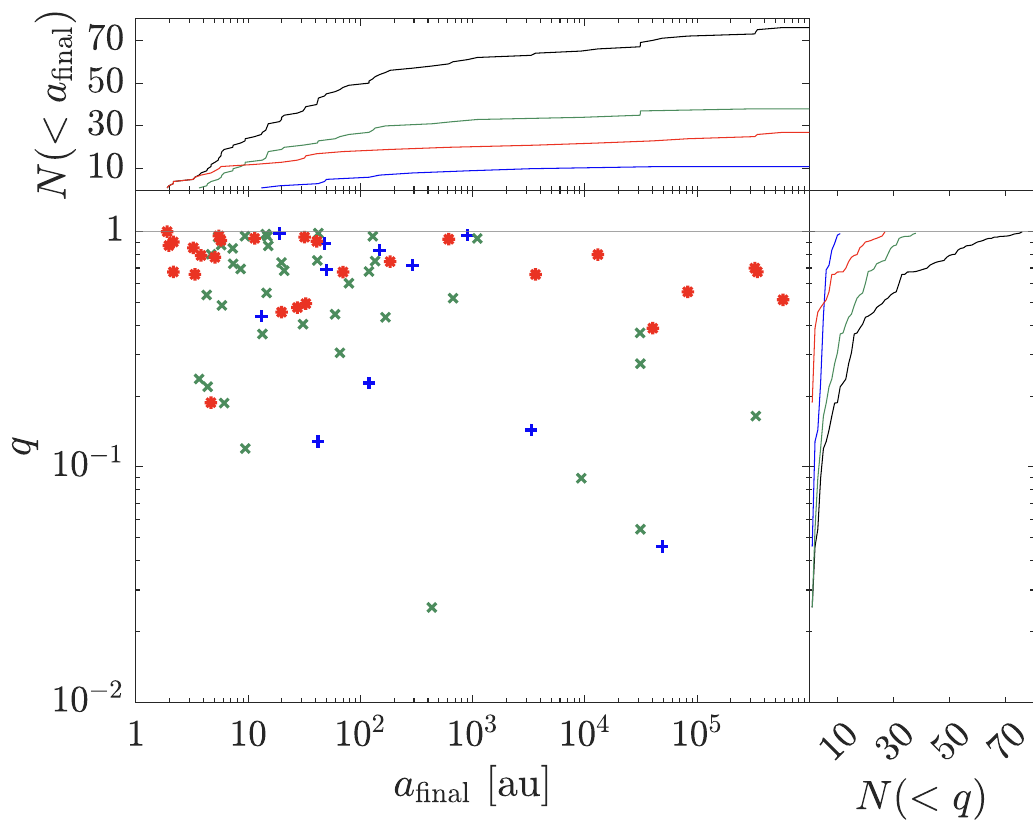}
    \caption{
    Final symmetric mass ratio ($q:=M_{*,2}/M_{*,1}$, where $M_{*,1}$ is the most massive star in the binary) versus final separation ($a_{\rm final}$).
    The horizontal dashed grey line indicates a mass ratio of unity.
    The colour scheme for the masses is the same as in Fig. \ref{fig:ainit_afinal}.
    The cumulative number of systems is shown along the top and right axes.    
    }
    \label{fig:q_afinal}
\end{figure}

\begin{figure}
    \centering
    \includegraphics[width=\hsize,trim={1.5cm 7.5cm 2.5cm 8.25cm},clip]{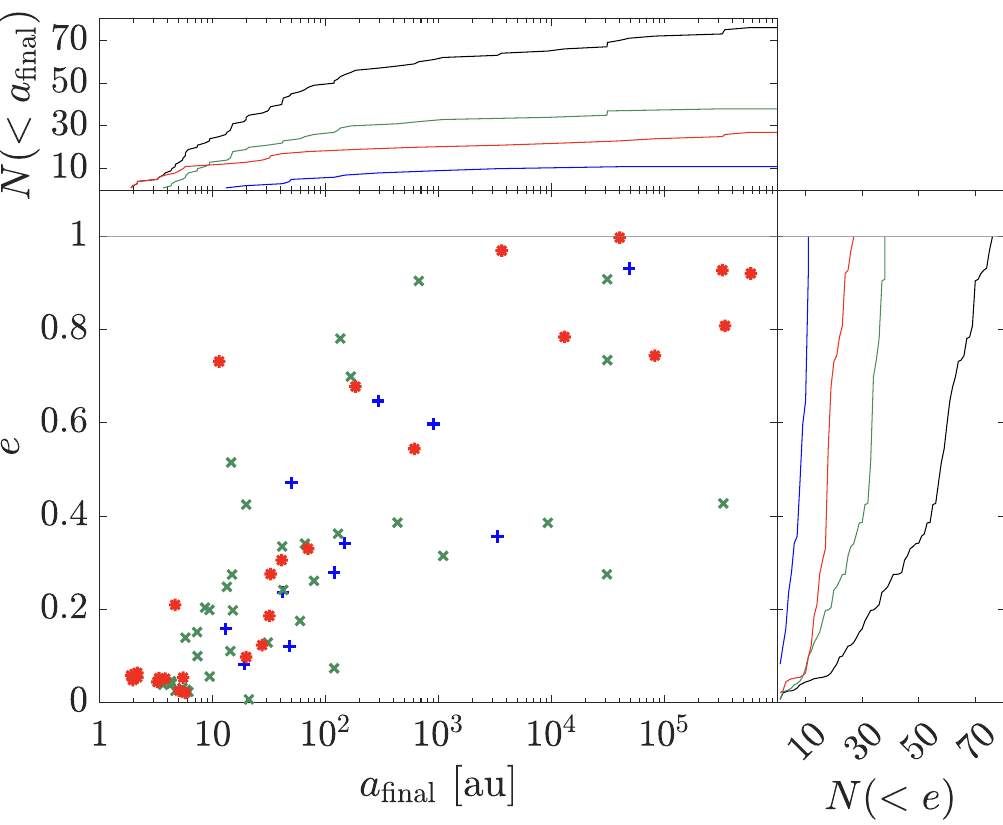}
    \caption{
    Eccentricity versus separation of binary systems at the end of the simulation. 
    The colour scheme for the masses is the same as in Fig. \ref{fig:ainit_afinal}.
    The cumulative number of systems is shown along the top and right axes.
    }
    \label{fig:ecc_afinal}
\end{figure}

\subsection{Migration}
In this Section, we focus on the migration of binary systems, particularly the time evolution of their separations.

Figure~\ref{fig:separation_evolution} shows the time evolution of binary separations for systems with primaries more massive than $2~M_\odot$, where we define the primary as the most massive star in the isolated- or inner-binary system.
The black and red lines trace the binary separations, with red segments marking periods when the binary interacts with a circumstellar or circumbinary disc.
We find that the binary separation remains close to its initial value for the first $\sim10~\mathrm{kyr}$, during which the stars gradually approach one another.
Following this phase, interactions with the gas in circumstellar discs lead to a sudden decrease in separation, typically to $10$–$1000~\mathrm{au}$.
In some systems, the separation continues to shrink after disc interaction, eventually reaching values below $10~\mathrm{au}$ as a result of interactions with the circumbinary disc.
Here we do not distinguish between circumstellar and circumbinary discs; the circumbinary interaction phase is considered part of the overall disc-interaction phase.

Massive systems frequently undergo dynamical interactions with nearby stars or gas clumps. The green lines in Fig.~\ref{fig:separation_evolution} show the evolution of the separation between the primary star and a third body, while the grey lines indicate epochs when one of the binary members merges with another star. Since massive stellar systems tend to form in turbulent cores (Figs.~\ref{fig:binary_fig_individual} and~\ref{fig:fragmentation_mass}), they are more likely to experience close interactions or mergers with neighboring objects.
This introduces stochasticity into the evolution of the binary separation, meaning that the separation does not necessarily decrease monotonically over time even when the binary interacts with the disc.

Figure~\ref{fig:mass_accretion_analysis} shows the time evolution of the stellar masses in systems where the primary has a mass greater than $2~M_\odot$.
The black, blue, and green lines represent the mass evolution of the primary, secondary, and outer tertiary stars, respectively. We find that the mass ratio between the primary and secondary stars often tends to approach unity during the course of the binary evolution, whereas the mass of the outer tertiary star evolves independently of the inner binary.

In Section~\ref{sec:isolated_binaries}, we first examine isolated binaries to present the general picture of binary evolution. 
We then turn to more complex multi-body interactions in Section~\ref{sec:complex_binaries}.
Finally, we present the statistical properties of binaries in Section~\ref{subsubsec:results:statistics}.

\subsubsection{Clean assembly} \label{sec:isolated_binaries}
Figure \ref{fig:circum_binary_fig} illustrates the formation of three representative systems.
These systems span different mass regimes, with intermediate- or high-mass primaries, their stellar components do not experience stellar mergers and undergo relatively few dynamical interactions with other stars. 
Two of them, ID 2 and ID 37, ultimately assemble an isolated binary with no tertiary companions and no stellar mergers after binary formation (top and bottom panels in Fig.~\ref{fig:circum_binary_fig}, respectively).
The middle panel (ID 18) presents the most massive binary in our sample with the final separation smaller than $10~$au, with an inner primary mass of $5.9~M_\odot$.
This sample is relatively isolated in a sense that no stellar member experiences the stellar mergers.
Such isolated systems are common among binaries with intermediate- and low-mass primaries ($<6~M_\odot$), but they are rare in massive binaries in our simulations, with only one example found out of 18 systems.

The evolution of these binaries can be broadly divided into three phases: initial contraction, disc migration, and the circumbinary phase.
We define the circumbinary phase as the stage in which each circumstellar disc becomes fully embedded within the disc of its companion star.
The left panel of the figure shows the time evolution of the binary separation, while panels A–D display the projected gas density around the system at key moments: the formation of the binary members (A), and the subsequent three evolutionary phases (B-D).

During the initial contraction phase, the binary separation decreases due to gravitational collapse of the progenitor cloud core or through few-body interactions among fragments formed within the core. 
The binary member stars approach each other along dense filaments (e.g., ID 29 in Fig.~\ref{fig:binary_fig_individual}), spiral arms 
(ID 2 in Fig.~\ref{fig:circum_binary_fig}), or in a more stochastic manner via interactions with other clumps (IDs 18 and 37). 
The typical timescale for the initial contraction corresponds to the dynamical time of the progenitor cloud, which is approximately $10$--$100~\mathrm{kyr}$ (Eq.~\ref{eq:t_dynamical}). 
The initial binary orbits are generally eccentric, and the binary separation exhibits oscillatory behaviour over time. The eccentricity gradually decreases, as indicated by the decline in apocentric separation, that is, the peak of the separation decreases with time. This reduction is driven by dynamical interactions with nearby stars or by tidal interactions with surrounding discs.
Tidal interaction is expected to play a significant role, as the typical pericentric distance during the first close passage is on the order of several hundred au—only a factor of a few larger than the size of circumstellar discs. At these scales, tidal forces between the discs become important for promoting inward migration \citep{1994ApJ...421..651A, 2018MNRAS.475.5618B}.

After the initial contraction phase, the separation becomes smaller than the size of the circumstellar disc ($\sim100$ au), and the binary components begin interacting with the disc gas (panels C in Fig.~\ref{fig:circum_binary_fig}). 
Spiral arms rapidly extract angular momentum from the binary system, reducing the separation to a few $10$ -- $100~\mathrm{au}$.  
This disc migration phase ends via one of the following two mechanisms: 1) the disc gas is either accreted or expelled, preventing further angular momentum transport; or 2) the gas mass within the binary orbit becomes significantly lower than the binary mass, resulting in the system being surrounded by a circumbinary disc.

Tight binary systems with separations smaller than several tens of au further evolve through a circumbinary disc phase. In this stage, the entire binary system is embedded within a circumbinary disc, and the separation gradually shrinks to below $10~\mathrm{au}$. 
All binaries that reach final separations smaller than $10~\mathrm{au}$ undergo this 
circumbinary phase.
The structure of the circumbinary disc is more stable and less perturbed compared to that of circumstellar discs (panels D in Fig.~\ref{fig:circum_binary_fig}), while a small non-axisymmetric spiral arm can be seen in ID 18. 
During this phase, the binary separation continues to decrease, although the migration timescale is significantly longer than that in the earlier disc interaction phase. 
Continuous mass accretion onto the system sustains the circumbinary disc, and binary-disc interaction efficiently extract angular momentum, leading to the formation of tight binaries with separations below $10~\mathrm{au}$. 
This phase ends when radiative feedback disperses the gas in the circumbinary disc. In the case of massive binaries, radiation from the massive stars themselves is sufficient to clear out the disc gas (e.g., ID 2). For low-mass binaries, the stellar radiation is insufficient to do so, and instead, 
external irradiation from nearby massive stars photoevaporates the circumbinary disc, effectively halting further migration.

\begin{figure}
    \centering
    \includegraphics[width=\hsize]{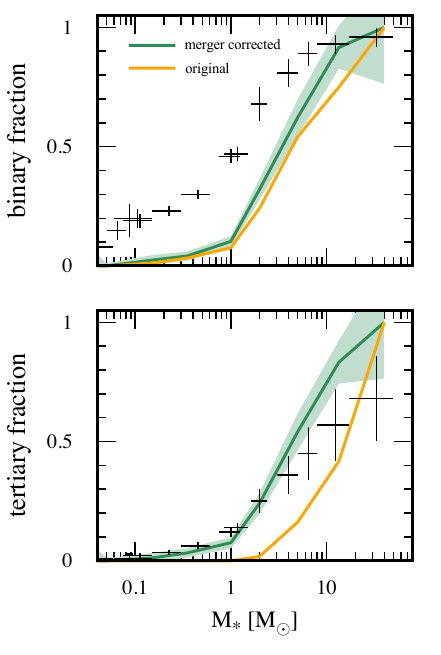}
    \caption{
    Binary (top panel) and tertiary (bottom panel) fraction as a function of the primary mass. 
    For this particular analysis and plot, the primary mass is defined as the most massive star in the system. In triple systems, it could be either a component of the inner binary or the tertiary star.
    The yellow (or green) lines show the fraction using the original (or merger-corrected) data (Section~\ref{sec:merger_correction}). 
    The green shaded regions are 1$\sigma$ error for the binary and tertiary fractions for merger-corrected samples. 
    The black crosses are the observed binary and tertiary fractions taken from \citet{2023ASPC..534..275O}.
    }
    \label{fig:multiplicity}
\end{figure}

\subsubsection{Complex assembly} \label{sec:complex_binaries}
Figure~\ref{fig:merged_binary_fig} shows the evolution of the final three-body system, highlighting its multi-body interactions and mergers with other stars.
We focus on systems ID 0 and ID 1, which host the most and second-most massive primaries in our simulations. Panel A shows the density distribution at the birth of the multiple system, originating from the fragmentation of a dense filament.
These fragments collapse to form protostars, which approach one another during the initial contraction phase (panel B).
In panel C, stars interact with the surrounding circumstellar disc, while the central stars have already formed a binary companion.
Through continued dynamical interactions between the stars and the initial central binary, together with disc–gas interactions, the system evolves into a tight binary surrounded by a circumbinary disc (panel D). The central tight binary then interacts with an outer third or fourth body, as well as with the gas supplied by the circumbinary disc, ultimately leading to stellar mergers.

At the time of the stellar merger shown in the figure, the final surviving binary companion is located several hundred au from the central binaries.
Although these stars later approach each other through gravitational interactions or scattering with other stars, strong ionizing feedback clears the surrounding gas beforehand.
As a result, the final binary members do not undergo a disc-interaction phase or enter a circumbinary-disc configuration, and their separations remain larger than $100~\mathrm{au}$.

\subsubsection{Statistics} \label{subsubsec:results:statistics}
Figure~\ref{fig:ainit_afinal} summarises the migration of binaries by comparing the separation of the binary at formation ($a_\text{initial}$) and to the separation of the binary in the final snapshot ($a_\text{final}$).

The top panel directly shows $a_\text{final}$ as a function of $a_\text{initial}$. 
The initial binary separations span a wide range, from $100$ to $10^6~\mathrm{au}$, with three agglomeration located at $a_\text{initial} \sim \{100,1000,3 \times 10^5\}$ au. 
These regions correspond to different binary formation channels: the first region ($\sim 100~\mathrm{au}$) is primarily associated with disc fragmentation, the second ($\sim 1000~\mathrm{au}$) is dominated by filament and core fragmentation, and the third ($\sim 3 \times 10^5~\mathrm{au}$) corresponds to dynamical capture.
Despite the broad distribution in initial separations, the final binary separations tend to converge to a few to several tens of au. A clear transition is observed at a critical initial separation $a_\text{crit} \sim 2 \times 10^4~\mathrm{au}$. 
For systems with $a_\text{initial} > a_\text{crit}$, the binary separation remains largely unchanged throughout the evolution. 
In contrast, for binaries with $a_\text{initial} < a_\text{crit}$, more than two-thirds undergo a reduction in separation by more than an order of magnitude. 

The origin of the critical separation is closely related to the characteristic fragmentation scales. 
The fragmentation scale is roughly set by the Jeans length of the initial filament \citep{2001ApJ...559L.149I}, which can be estimated as
\begin{align}
a_{\mathrm{crit}} \sim l_{\mathrm{J}} 
= 3.5\times 10^4~\mathrm{au} 
\left( \frac{T}{10~\mathrm{K}} \right)^{1/2}
\left( \frac{\mu}{2.2} \right)^{-1/2}
\left( \frac{n}{10^4~\mathrm{cm^{-3}}} \right)^{-1/2},
\end{align}
where $c_{\mathrm{s}}$ is the sound speed and $\mu$ is the mean molecular weight.
If a pair of stars forms within a fragmented core, they interact dynamically and their separation decreases during the gravitational collapse of the core.
The initial orbital energy is removed through interactions with the filament or disc gas, or through stochastic multi-body encounters.
In contrast, if the separation is larger than the typical size of the fragmented cores, the system does not necessarily undergo orbital shrinkage; instead, a bound binary may simply form through random encounters.
In this case, the initial orbital energy is difficult to dissipate, and the binary separation remains close to its initial value.

The bottom panel of Fig.~\ref{fig:ainit_afinal} shows the ratio between the final and initial separations, $a_\text{final} / a_\text{initial}$, as a function of the initial binary separation. 
When $a_\text{initial}$ lies between $200$ and $2 \times 10^4~\mathrm{au}$, the values of $a_\text{final} / a_\text{initial}$ are clearly divided into two distinct populations: either close to $1$ or below $0.01$. 
Notably, there are no binaries with $a_\text{final} / a_\text{initial} \sim 0.1$.
This bimodality indicates that once migration becomes effective, the binary separation rapidly shrinks by more than an order of magnitude, rather than evolving gradually.

Figure~\ref{fig:q_afinal} shows the symmetric mass ratio of the binary system, $q \equiv M_{*,2} / M_{*,1}$, where $M_{*,1}$ is the more massive star in the binary, as a function of the final binary separation $a_\text{final}$. 
As $a_\text{final}$ decreases, the mass ratio distribution becomes increasingly peaked toward unity. 
This trend suggests that, during the inward migration, the accreting gas is shared more equally between the two components, favouring the formation of equal-mass binaries.

Figure~\ref{fig:ecc_afinal} shows the orbital eccentricities of binary systems as a function of their final separation. 
A clear trend is observed: the eccentricity decreases with decreasing $a_\text{final}$. 
A rapid decline in eccentricity is seen below $\sim 100~\mathrm{au}$.
At large separations ($a_\text{final} \gtrsim 100$--$1000~\mathrm{au}$), the binary orbits retain dynamical imprints of their formation processes. 
Turbulence within the binary progenitor core, fragmentation along filaments, and dynamical interactions with nearby stars naturally lead to eccentric orbits.
As the separation decreases to $\sim 100~\mathrm{au}$, disc-star interactions become dominant and act to circularize the binary orbit. 
For systems with $a_\text{final} \lesssim 10~\mathrm{au}$, additional circularization is driven by interactions with the circumbinary disc.

\subsection{Multiplicity}
Figure~\ref{fig:multiplicity} shows the binary and tertiary fractions as a function of primary mass, demonstrating a strong mass dependence in stellar multiplicity. This trend is well established observationally and is reproduced in our simulations.

The high multiplicity fraction ($\gtrsim 90\%$) for primary stars with $M_* \gtrsim 8~M_\odot$ reflects their typical formation in dense environments, where disc and core fragmentation efficiently produce multiple bound companions. Moreover, stars formed in such dense regions have a higher probability of interacting with neighbouring stars, leading to the formation of binary or three-body systems.
In contrast, the low multiplicity observed at $M_* \lesssim 2~M_\odot$ arises because fragmentation occurs less frequently in circumstellar discs around low-mass stars. Our simulated binary fractions in this regime are significantly smaller than the observed values.

To account for unresolved binaries, which in our simulation are assumed to merge but may in reality survive as tight binaries with separations smaller than $\sim 1~\mathrm{au}$, we apply a correction.
The green lines show the binary and tertiary fractions under the assumption that all merged protostellar pairs instead survive as unresolved tight binaries.
Even with this correction, however, the binary fraction for low-mass systems remains below the observed values. This discrepancy may indicate that additional physical processes, such as magnetic braking or radiative feedback, or higher numerical resolution, are required to fully capture the multiplicity in this mass range (Section~\ref{sec:discussion_multiplicity}).

\subsection{Mergers}
In binary and higher-order multiple-star systems, stellar mergers play a crucial role in decreasing the number of bound objects and can drastically affect the evolution of the merger remnant \citep[e.g.,][for a recent review]{2025arXiv250918421S}. In star-forming regions, binary formation and mergers frequently occur together, driven primarily by the complex dynamics during star formation rather than by standard stellar or binary evolution processes.

\begin{figure}
    \centering
    \includegraphics[trim={1.5cm 2cm 1.5cm 0},clip,width=\columnwidth]{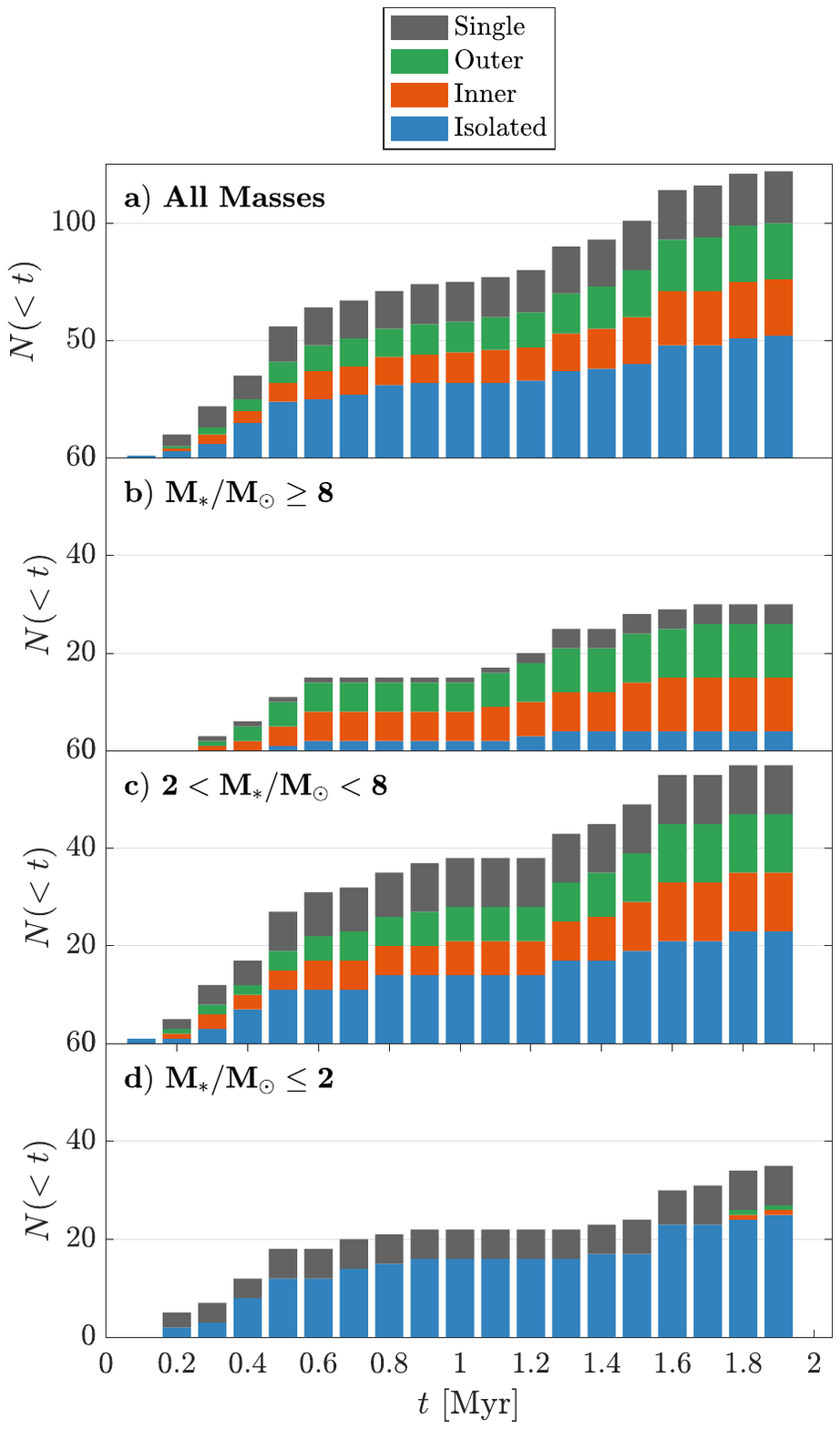}     
    \caption{
    Cumulative number of binaries formed over time. 
    We distinguish between isolated binaries (blue), binaries that form part of triple systems, composed of an inner (red) and an outer (green) binary, and binaries that merge and lead to a single remnant (grey).
    Panel a shows all binaries in our simulation.
    Panel b shows systems in which the most massive star is $M_*/M_{\odot} > 8$.
    Panel c shows systems in which the most massive star is $2 < M_*/M_{\odot} < 8$.
    Panel d shows systems in which the most massive star is $M_*/M_{\odot} < 2$.
    Note that the y-axis scales for panel a differ from those in panels b–d.
    }
    \label{fig:time_of_binary_formation_CDF}
\end{figure}

Figure \ref{fig:time_of_binary_formation_CDF}  illustrates how the number of binaries evolves over time. Binary formation follows the star formation history \citep[cf. Figure 2 of][]{2024MNRAS.530.2453C}, resulting in 52 binaries, 24 triples, and 22 binaries that eventually merge into single stars. There is a direct relationship between mass and multiplicity: only one triple is found in the low-mass star system (panel d), compared to 12 and 11 triples in intermediate-mass (panel c) and high-mass (panel b) star systems, respectively.

\begin{figure}
    \centering
    \includegraphics[trim={1.5cm 2cm 1.5cm 0},clip,width=\columnwidth]{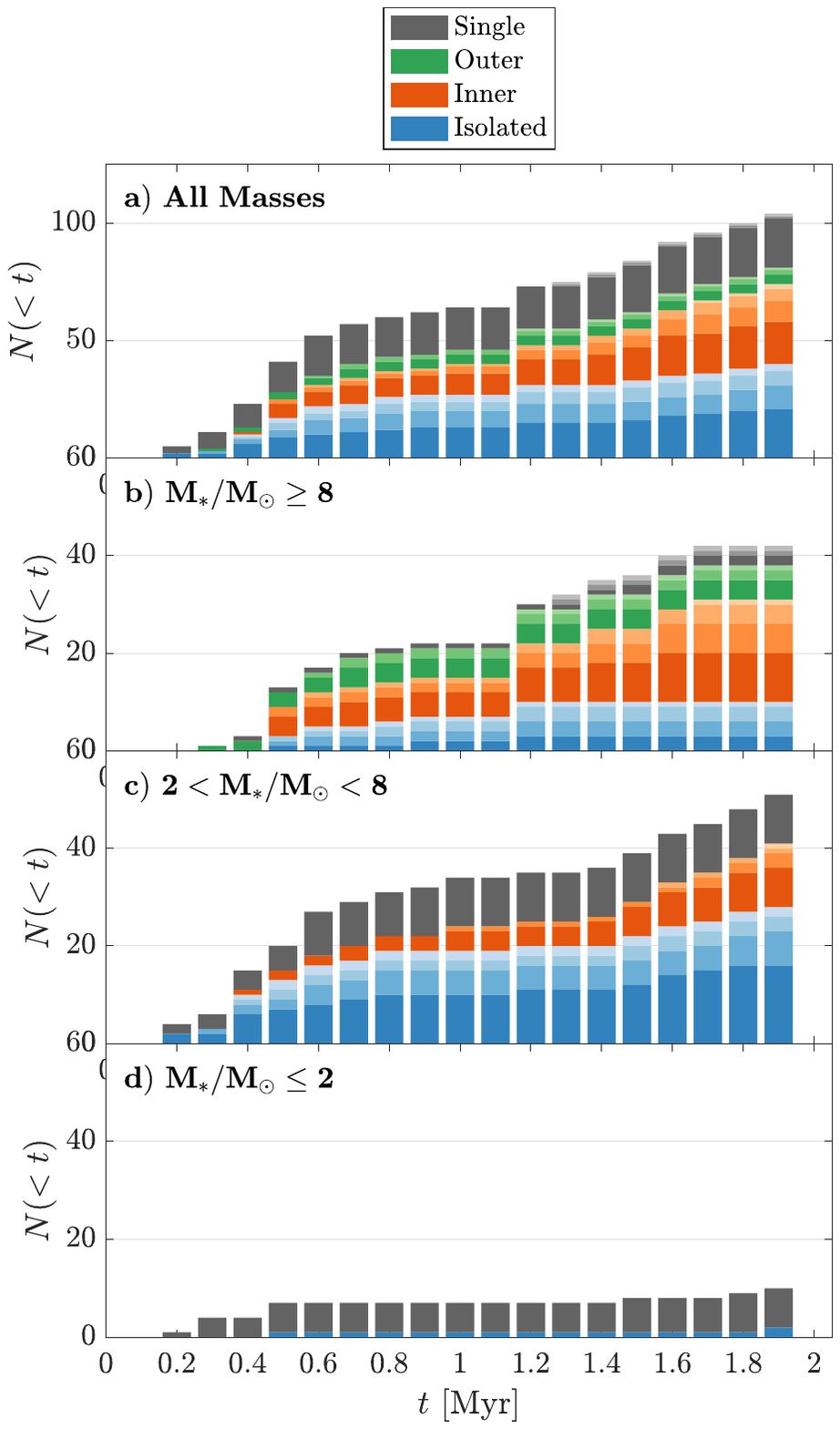}      
    \caption{
    Cumulative number of merger events over time. 
    The colours and panels are the same as in Fig. \ref{fig:time_of_binary_formation_CDF}.
    For each colour, the shading depicts the number of merger events experienced per binary: one (opaque), two (dark), three (light), and even four (very light).
    }
    \label{fig:time_of_merger_events_CDF}
\end{figure}

Figure \ref{fig:time_of_merger_events_CDF} shows the evolution of merger events, which also correlates with mass and multiplicity. Not only do intermediate-mass (panel c) and high-mass (panel b) star systems experience a greater number of merger events, but the most massive systems also undergo up to four merger events. In triple systems, merger events in the outer tertiary occur exclusively in the most massive star systems.

\begin{figure*}
    \centering
    \includegraphics[width=\hsize]{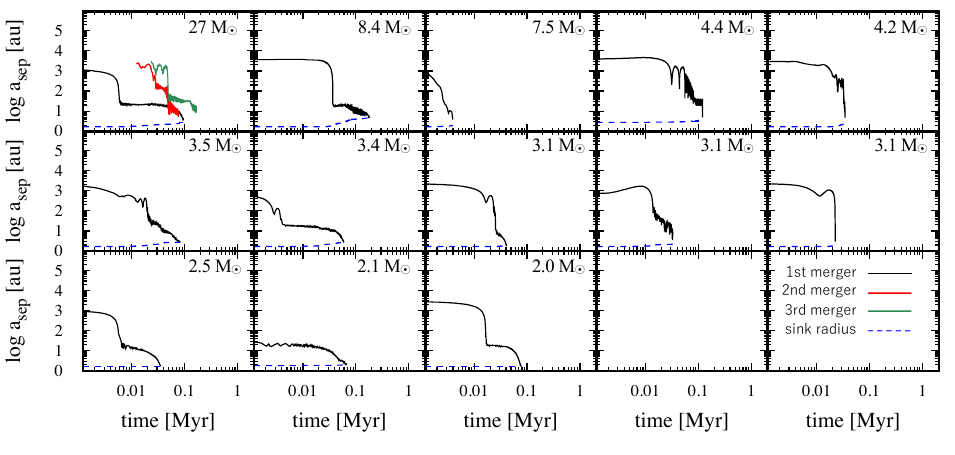}
    \caption{
    Time evolution of the orbital separation ($a_{\rm sep}$) for binary systems that eventually merge and survive as single isolated stars by the end of the simulation. 
    Time is measured from the formation epochs of the binary member stars that eventually merge. 
    The spatial resolution of the binary system—calculated as the sum of the sink radii of the two stars—is indicated by a dashed blue line. 
    The top-left panel shows the most massive merged system, which is the only case that undergoes multiple merger events; the second and third merger events are shown by red and green lines, respectively.
    }
    \label{fig:separation_evolution_merged}
\end{figure*}

Figure \ref{fig:separation_evolution_merged} illustrates the separation evolution of 13 binaries that merge, eventually becoming single stars. The most massive star in this sample experiences three consecutive mergers:
i) at 1.35 Myr (0.1 Myr after the formation of the initial binary consisting of equal-mass $1.4+1.4\ M_{\odot}$ stars), the first merger occurs, ii) at 1.38 Myr, a second merger takes place involving the first merger remnant—which has grown to $3.9\ M_{\odot}$—and a $3.0\ M_{\odot}$ companion, and iii) finally, at 1.5 Myr, a merger between the second remnant (now grown to $14.3\ M_{\odot}$) and a $5.9\ M_{\odot}$ star occurs. By the end of the simulation (2 Myr), this remnant has grown to $27.7\ M_{\odot}$, and is one of the most massive stars in our simulation.
Merger events can have significant implications for stellar properties, such as chemical composition, structure, and age determination, particularly in massive star systems; these are discussed in Section~\ref{subsec:disc:mergers}.


\subsection{Inclinations}
The orientation of a binary orbit may encode valuable information about the system's origins and influence its future evolution. 
Here, we analyse this orientation by examining two types of inclinations.

\begin{figure}
    \centering
    \includegraphics[trim={1.25cm 9.5cm 2cm 10cm}, clip, width=\hsize]{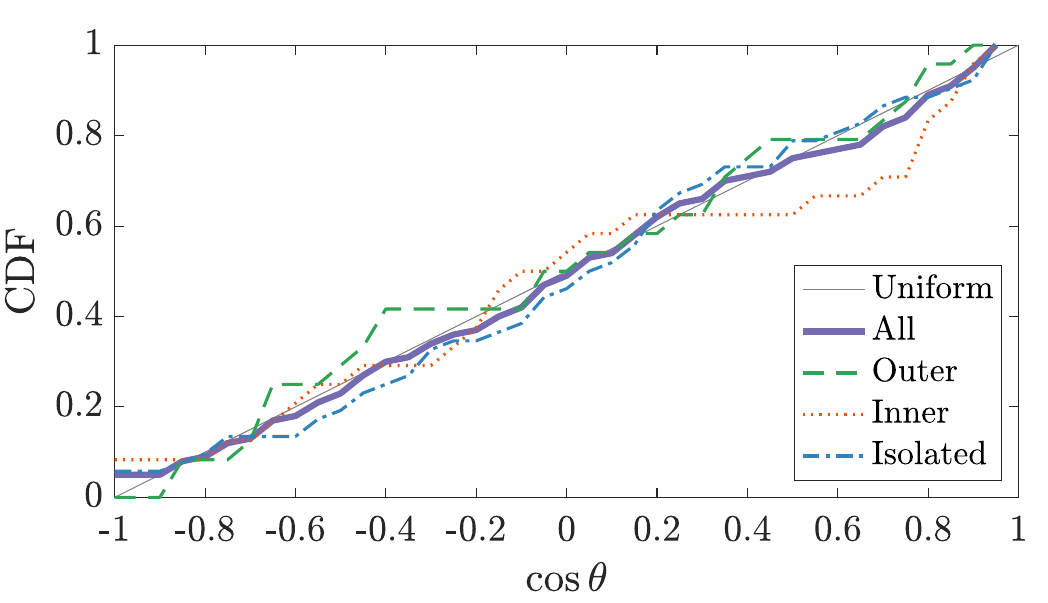}    
    \caption{
    Inclination angle ($\theta$) between the initial rotation axis and the final orbital angular momentum vector, represented as a cumulative distribution function (CDF).
    We show the analytical solution for a uniform distribution in $\cos\theta$ (thin solid black line) and the simulation results for all binaries (thick solid purple line). 
    Additionally, we present specific subsets from our simulation: outer binaries (dashed green line), inner binaries (dotted red line), and isolated binaries (dash-dotted blue line).
    }
    \label{fig:inclination}
\end{figure}

Figure~\ref{fig:inclination} displays the cumulative distribution function of the binary inclination ($\theta$) relative to the initial rotation axis of the star-forming cloud. 
The thick solid purple line represents the combined distribution of inner, outer, and isolated binaries as a function of $\cos \theta$.
This distribution closely follows the isotropic (i.e., uniform in $\cos \theta$) distribution, represented by the thin grey line, with a small excess of both aligned and anti-aligned systems ($\cos \theta = \pm 1$).
The green dashed, red dotted, and blue dash-dotted lines, which correspond to outer, inner, and isolated binaries respectively, are also consistent with isotropy. This isotropy likely originates from the turbulent motion dominating the initial cloud dynamics, which imposes locally random orientations on the forming stars. Moreover, dynamic interactions during binary assembly may reinforce—or at least preserve—this inherent randomness.
In the initial condition of the cloud, we have imposed a rigid rotation of $0.1\%$ of the gravitational energy. The rotation motion has $1000$ times smaller energy compared to the turbulent motion. 

\begin{figure}
    \centering
    \includegraphics[trim={1.25cm 9.5cm 2cm 10cm}, clip, width=\hsize]{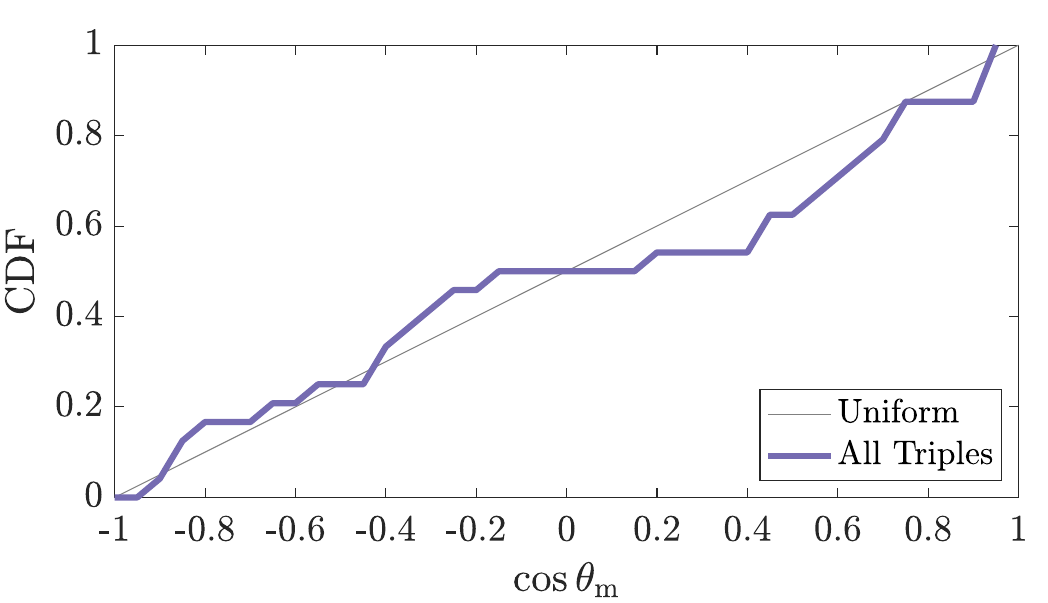}    
    \caption{
    Mutual inclination angle ($\theta_{\rm m}$) between the orbital angular momentum vectors of the inner and the outer of orbits in triple systems, represented as a cumulative distribution function (CDF).
    We show the analytical solution for a uniform distribution in $\cos\theta_{\rm m}$ (thin solid black line) and the simulation results for all triples (thick solid purple line).   
    }
    \label{fig:inclination_mutual}
\end{figure}

Figure~\ref{fig:inclination_mutual} shows the cumulative distribution of the angle between the binary orbits ($\theta_\text{m}$) for the inner and outer binaries in triple systems.
The thick solid purple line shows the distribution of all triples as a function of $\cos \theta_\text{m}$.
This distribution also closely follows the isotropic distribution, with small deficit (or excess) of anti-aligned (or aligned) binaries.
This isotropy most likely stems from the complex dynamics involved in forming multiple star systems (Figure \ref{fig:separation_evolution}). Because most of these triples belong to high-mass systems—which tend to form near other massive stars—they frequently undergo star mergers, exchanges, dynamical instabilities, or a combination of these processes.

\begin{figure}
    \centering
    \includegraphics[width=\hsize]{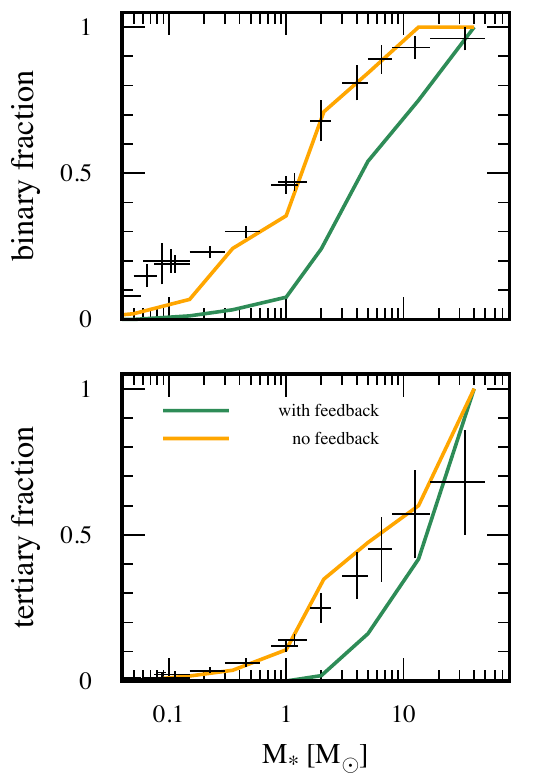}
    \caption{
    Binary (top panel) and tertiary (bottom panel) fraction as a function of the primary mass. 
    The green and yellow lines show the fraction using the data from the base simulation---which includes feedback---and alternative run without feedback, respectively. 
    The black crosses are the observed binary and tertiary fractions taken from \citet{2023ASPC..534..275O}.
    }
    \label{fig:multiplicity_comp}
\end{figure}

\section{Discussion}\label{sec:discussion}
\subsection{Fragmentation}
We have presented the statistics of binary and higher-order multiple systems, focusing on their formation pathways. Now, we discuss how fragmentation at different spatial scales contributes to the origin of binaries and determines their separations.
As shown in Fig.~\ref{fig:overall}, our simulated molecular cloud exhibits fragmentation across a hierarchy of spatial scales. The large-scale filamentary structure of $\sim1-10 $ pc initially fragments into dense cores with typical sizes of $\sim 0.1~\mathrm{pc}$. At smaller scales, further fragmentation occurs within these cores and along denser filamentary substructures with characteristic widths of $\sim 10^3~\mathrm{au}$. Finally, disc fragmentation is observed at scales below $\sim 100~\mathrm{au}$. 
This stepwise, multi-scale fragmentation pattern is consistent with recent high-resolution (sub-)millimetre observations, which reveal hierarchical fragmentation from parsec to sub-100 au scales \citep[e.g.,][]{2018ApJ...853....5P,2025ApJ...994..233Y}. 
Notably, ALMA observations of nearby star-forming regions have directly identified core-scale fragmentation with separations of several $10^3$--$10^4~\mathrm{au}$, supporting the idea that this intermediate-scale fragmentation mode plays a significant role in setting the initial conditions for binary formation \citep[e.g.,][]{2023ApJ...950..148M, 2024ApJ...966..171M,2024NatAs...8..472L,2025ApJ...994..233Y}.
These observations support the view that multiple fragmentation mechanisms operate at different stages of collapse, contributing to the origin and diversity of stellar multiplicity.

The initial separation of binary systems depends on the fragmentation mode and exhibits a multi-modal distribution, with prominent peaks at $\sim100~\mathrm{au}$ and $\sim10^3~\mathrm{au}$, corresponding to disc fragmentation and core/filament fragmentation, respectively (distributions in top panel of Fig.~\ref{fig:ainit_afinal}). However, this multi-modality is erased during the subsequent evolution. As shown in the right panels of Fig.~\ref{fig:ainit_afinal}, the distribution of final binary separations develops is more smooth, with accumulating systems at $\lesssim 10~\mathrm{au}$.
This trend is in excellent agreement with recent high-resolution radio observations of Class 0 and Class I protostars in the Perseus \citep{2016ApJ...818...73T} and Orion \citep{2022ApJ...925...39T} clusters. These studies report that Class 0 binaries exhibit a bimodal separation distribution with peaks at $\sim 100$ and $10^3~\mathrm{au}$, whereas such bimodality disappears in Class I systems. Our simulations naturally reproduce this observed evolution as a consequence of dynamical migration, whereby companions move from wide to tight separations.
The typical migration timescale found in our simulation is $\sim 0.1~\mathrm{Myr}$, broadly consistent with the observationally inferred timescale of $0.2$--$0.5~\mathrm{Myr}$ reported by \citet{2022ApJ...925...39T}.

Our simulations reveal that most tight binaries with separations smaller than $100~\mathrm{au}$ originate from filament and core fragmentation. In contrast, only about $\sim 20\%$ of such systems form via disc fragmentation, challenging the common theoretical hypothesis that tight binaries predominantly arise from disc fragmentation \citep[e.g][]{2016ARA&A..54..271K}. 
However, our findings do not exclude the contribution of disc fragmentation entirely. Disc fragmentation is observed in both simulations \citep{2010ApJ...708.1585K} and high-resolution observations \citep{2016Natur.538..483T}, although it may represent a subdominant formation channel. 
This conclusion is supported by recent observations of high-mass star-forming regions by \citet{2024NatAs...8..472L}, who report fragmentation of a $\sim 0.1~\mathrm{pc}$ cloud into 44 protostellar systems, among which 20 exhibit multiplicity. The observed binary separations range from $200$ to $1400~\mathrm{au}$, and no evidence of disc-like rotation is found. This strongly suggests that these systems originate from core fragmentation rather than disc fragmentation.

\subsection{Migration}
\subsubsection{Origin of close binaries}
We identify two distinct phases of binary hardening.
The first occurs immediately after binary formation and is likely associated with the same mechanisms that drive star formation; it operates over $\sim10$–$100$ kyr.
Cloud collapse and few-body interactions within the progenitor core reduce the binary separation to $\sim100$–$1000~\mathrm{au}$.
The second phase occurs in the presence of a disc and is driven by disc–binary interactions as the disc transitions from thick to thin, typically over $\gtrsim0.1$ Myr, consistent with the model presented in \cite{2020MNRAS.491.5158T}.
Close systems may therefore begin forming very early in the star-formation process—during the optically thick phase or within rapidly expanding HII regions that later become optically thin \citep[e.g.,][]{2025arXiv250918322W}.

At our current resolution, we cannot resolve tidal dissipation, which may dominate the final stages of hardening.
Recent observational studies \citep{2021A&A...645L..10R,2024A&A...690A.178R} of clusters at different ages suggest that massive, close ($\sim$1 d) binaries undergo significant hardening within the first 1–2 Myr after formation.
The separations involved in these observed systems are far tighter than those reached in our simulation.
They concluded that the tidal processes are the most likely drivers on Myr timescales among the several other shrinking mechanisms. 

Binary-disc interactions may also be crucial for migration, enabling the binary to migrate in or out by a fraction approximately equal to the accreted mass per binary mass \citep[][i.e., $\dot M / M \sim \dot a / a$]{2024A&A...688A.128V}.
Recent high-resolution hydrodynamics simulations broadly agree that equal-mass, circular binaries in relatively warm pressure-supported discs, with typical aspect ratio $\sim 0.1$, yield binary outspiral \citep[][and references therein]{2024ApJ...970..156D}.
However, similar simulations shown that the sign of binary migration can vary according to many parameters like binary eccentricity \citep{2021ApJ...909L..13Z, 2021ApJ...914L..21D}, mass ratio \citep{2020ApJ...901...25D, 2024ApJ...967...12D}, and disc thermodynamics/aspect ratio \citep{2020ApJ...900...43T}.
Binary outspiral, though, is generally a property of warmer and thicker discs where the binary accretes efficiently \citep{2017MNRAS.466.1170M, 2019ApJ...871...84M}, while inspiral is ubiquitous for thinner, cooler systems \citep{2025ApJ...984..144T}.

Our results should therefore be regarded as complementary to idealised simulations with tidal effects and circumbinary-disc calculations, capturing the early stages of binary hardening in a turbulent, clustered star-forming environment.

\subsubsection{Comparison to observations}
In our simulation, the final binary separation distribution peaks around $\sim 10~\mathrm{au}$, consistent with observational results for stars with $M_* \lesssim 8\ M_\odot$ \citep{2010ApJS..190....1R, 2019AJ....157..216W, 2023ASPC..534..275O}. 
In contrast, in our simulation we do not observe a prominent peak in the final separation distribution for stars more massive than $M_*\gtrsim8\ M_{\odot} $, which would correspond to spectral type OB stars ($M_*\gtrsim2\ M_{\odot} $).
This may be attributed to merger prescription in our simulation. 
Many of these stars (8 out of 11 stars) undergo multiple mergers during their evolution. 
This suggests that, in these systems, at least one companion approached a separation of $1$--$10~\mathrm{au}$.
Given that actual stellar radii are only several $R_\odot$, much smaller than our sink radius, such close encounters may not necessarily result in physical coalescence. Instead, they could survive as tight binary systems, potentially populating the small-separation tail of the distribution.

We have also found a strong correlation between binary separation and both mass ratio and orbital eccentricity, as discussed in Section~\ref{subsubsec:results:statistics}. In particular, systems with separations $a > 100~\mathrm{au}$ exhibit a broad distribution of mass ratios, ranging from nearly equal-mass to extreme unequal-mass ($q\approx 0.03$) binaries. 
In contrast, closer systems with $a \lesssim 100~\mathrm{au}$ preferentially exhibit mass ratios near unity, which is in agreement with spectroscopic observations of high-mass stars \citep[e.g.,][]{2012MNRAS.424.1925C}. 
Similarly, orbital eccentricities tend to decrease with decreasing separation, with close binaries showing nearly circular orbits \citep[e.g.,][]{2017ApJS..230...15M}. 

These trends are likely driven by dynamical interactions between the stars and the surrounding circumstellar or circumbinary discs. When the binary separation is larger than the typical size of the circumstellar discs, the components evolve largely independently, leading to mass ratios that are essentially uncorrelated. Similarly, high eccentricities are expected in wide binaries, as their orbits are shaped by the global contraction of the natal filamentary structures or by stochastic multi-body interactions among protostars forming within the same core.
On the other hand, theoretical studies have suggested that sustained disc accretion tends to equalize the masses of binary components through differential accretion, thereby driving systems toward unity mass ratios at smaller separations \citep[e.g.,][]{2015MNRAS.452.3085Y,2017MNRAS.465..986S}. 

These observed distributions support the idea that gas dynamical processes play a crucial role in determining the final properties of binary systems, especially during the embedded accretion phase. 
Although our simulation qualitatively captures these trends, further work incorporating extended time baselines and additional physical processes (such as magnetic braking and radiative feedback) may be necessary to achieve closer quantitative agreement with observations.

\subsection{Multiplicity} \label{sec:discussion_multiplicity}

Our simulations reproduce a consistent binary fraction for massive stars but predict a lower fraction for lower-mass stars compared to local field populations (Fig.~\ref{fig:multiplicity}). We attribute this shortage of low-mass binaries to the strong effect of dust heating, which stabilizes circumstellar discs, dense filaments, and cores. This heating suppresses fragmentation on scales below $10^3$–$10^4\mathrm{au}$ \citep{2024MNRAS.530.2453C}. Indeed, when dust heating is not included, the resulting multiplicity distribution agrees well with observations (Fig.~\ref{fig:multiplicity_comp}). The suppression of fragmentation by radiative heating has also been emphasized in previous numerical studies \citep[e.g.,][]{2010ApJ...725.1485O, 2012MNRAS.419.3115B, 2017MNRAS.465....2M}. In particular, \citet{2010ApJ...725.1485O} showed that including dust heating substantially reduces the number of companions around low-mass stars. 

On the observational side, only a few studies have directly measured the binary or multiple star fractions in star-forming regions. For example, \citet{2018MNRAS.478.1825D} studied stars with masses between $0.8$ and $2.4\ M_\odot$ and found that while the binary fraction at separations greater than 60 au is significantly lower than that in the field, the incidence of closer binaries (with separations between 10 and 60 au) is higher than among field stars. This trend appears to conflict with our results, which show that local radiation strongly suppresses the binary fraction in low-mass stars.

The discrepancy may stem from our treatment of radiative transfer. Instead of explicitly solving the multi-wavelength radiation transport, we adopt a simplified spherical radiation model to describe the propagation of infrared photons responsible for dust heating effects \citep{2017JPhCS.837a2007F, 2018MNRAS.480..800H}. While this approximation reproduces the overall disc mid-plane temperatures reasonably well \citep{1999ApJ...525..330Y, 2018A&A...616A.101K, 2020MNRAS.497..829F}, it may artificially overheat the dense regions within spiral arms that would otherwise be shielded from stellar radiation.
\citet{2021MNRAS.502.3646G} compared simulations using this simplified treatment with those employing full multi-wavelength radiative transfer, and found that the differences in the resulting mass spectra were subdominant. 
\citet{2023MNRAS.518.4693G} have further argued that enhancing the background radiation field does not strongly impact the overall binary fraction.
However, it is important to note that their simulations did not resolve circumstellar discs, in which radiative shielding and heating may play a more critical role.

\subsection{Environmental Effects}
Our current analysis focuses on a single simulation at Solar metallicity. However, star formation spans a wide range of environments, and observations indicate that the stellar multiplicity—especially for massive stars—may be influenced by factors such as metallicity, turbulence, and gas density \citep[e.g.,][]{2021MNRAS.507.3593M, 2023ASPC..534..275O}.

For instance, low-metallicity gas, with typically reduced dust abundance and cooling efficiency, may suppress small-scale fragmentation and yield different binary properties \citep[e.g.,][]{2009MNRAS.399.1255M, 2019MNRAS.484.2341B, 2021MNRAS.508.4175C, 2023MNRAS.518.4693G, 2023MNRAS.526.3933M}. 
Observationally, there is an anti-correlation between metallicity and the close binary fraction\citep{2018ApJ...854..147B, 2019MNRAS.482L.139E, 2019ApJ...875...61M, 2025NatAs...9.1337S}.

Simulations have yet to converge on the effects of metallicity. 
Some studies report increased fragmentation and a higher rate of close binaries in low-metallicity environments owing to reduced dust opacity and consequent lower disc gas temperatures \citep{2019MNRAS.484.2341B, 2023MNRAS.526.3933M, 2025MNRAS.537..752B}. 
Conversely, other investigations find fewer close binaries, possibly due to decreased cooling and higher temperatures in circumstellar discs, filaments, and cores \citep{2021MNRAS.508.4175C, 2024MNRAS.530.2453C, 2023MNRAS.518.4693G}.

Although we have not yet analysed these effects in detail, our simulation suite includes runs with metallicities ranging from $10^{-4}$--$1~Z_\odot$. A systematic comparison of these runs will allow us to explore how metallicity shapes the formation and evolution of binary and multiple systems. This will be a key focus of future work aimed at understanding the environmental dependence of stellar multiplicity.

\subsection{Mergers}\label{subsec:disc:mergers}
Observations suggest that stellar mergers produce luminous transients, and their merger remnants are frequently linked to blue stragglers \citep[e.g.,][and references therein]{2026enap....2..449W} and blue supergiants \citep[e.g.,][]{2025MNRAS.542..703M}. 
Physically, mergers are prompt and occur preferentially in close binaries \citep[e.g.,][]{2012A&A...543A.126K}, leading to chemical abundances, structural changes, and rotation patterns that differ from those in undisturbed single-star systems \citep[e.g.,][for a recent review]{2025arXiv250918421S}.
\cite{2020MNRAS.491.5158T} suggest 33 per cent of the early-B primaries merged with a companion and \cite{2022NatAs...6..480W} propose that $\approx$20 per cent of blue main sequence stars in young open clusters are linked to merger events occurring early on the main sequence.

In our simulation, we tracked the orbital configurations of star particles involved in mergers. 
We found that mergers are common and more likely to occur in systems with intermediate-mass or high-mass stars. 
Among the intermediate-mass and high-mass star systems in our simulation (74 binaries and triples), the majority (63) include at least one star that has experienced a merger event. 
In high-mass star systems, 13 out of 15 systems contain a merger product.
We attribute this to a correlation between mass and multiplicity, which results from the formation of massive stars in denser, dynamically complex systems that are prone to close encounters and subsequent mergers (but see Section~\ref{subsubsec:disc:resolution} for a discussion on spatial resolution).

In our simulation, mergers play a crucial role in the growth of stellar mass. 
Compared to gas accretion—which typically occurs over longer timescales—mergers often contribute a comparable or even greater increase in mass (Figure \ref{fig:mass_accretion_analysis}). 
However, the theoretical outcome of these mergers is not yet fully understood, and modelling their remnants in our simulation is challenging due to their sporadic nature, repeated occurrences, and wide range of progenitor masses.

One notable prediction for stellar mergers is that they are the potential origin of magnetic massive stars \citep{2019Natur.574..211S}.
If we assume that every star that has undergone a merger becomes magnetic, $\approx$87\% of our high-mass star systems would host a magnetic star. 
In contrast, observations suggest that only about 10\% of stars in the upper main sequence are magnetic \citep{1992A&ARv...4...35L,2014IAUS..302..265W}, but recent estimates suggest that the fraction may reach 60\% for O-type stars \citep{2023MNRAS.521.6228H}. It is important to note that our estimate is crude, as we do not know how many mergers will actually yield a magnetic remnant, and some events may be artificially categorised as mergers due to resolution limitations (Section~\ref{subsubsec:disc:resolution}).

Stellar mergers could significantly affect the mass function \citep[e.g.,][for a recent review on the initial mass function]{2024ARA&A..62...63H}. If a correlation between mass, multiplicity, and stellar mergers exists, this relationship might influence the most massive star-forming regions in the Universe. For example, \cite{2018Sci...359...69S} report an excess of stars with masses greater than 30 $M_\odot$ in the 30 Doradus region of the Large Magellanic Cloud. We speculate that this excess may be partly due to stellar mergers like those discussed in this manuscript, suggesting that such mergers could be common in massive star-forming regions.
In contrast, \cite{2025ApJ...994..176M} report that the IMF slope for NGC 3603 is relatively normal; however, they note that forming massive stars is more challenging in the Milky Way---a higher-metallicity environment---compared to the LMC.

\subsection{Formation Times}
\begin{figure}
    \centering
    \includegraphics[trim={0.5cm 7.5cm 2cm 7.75cm}, clip, width=\hsize]{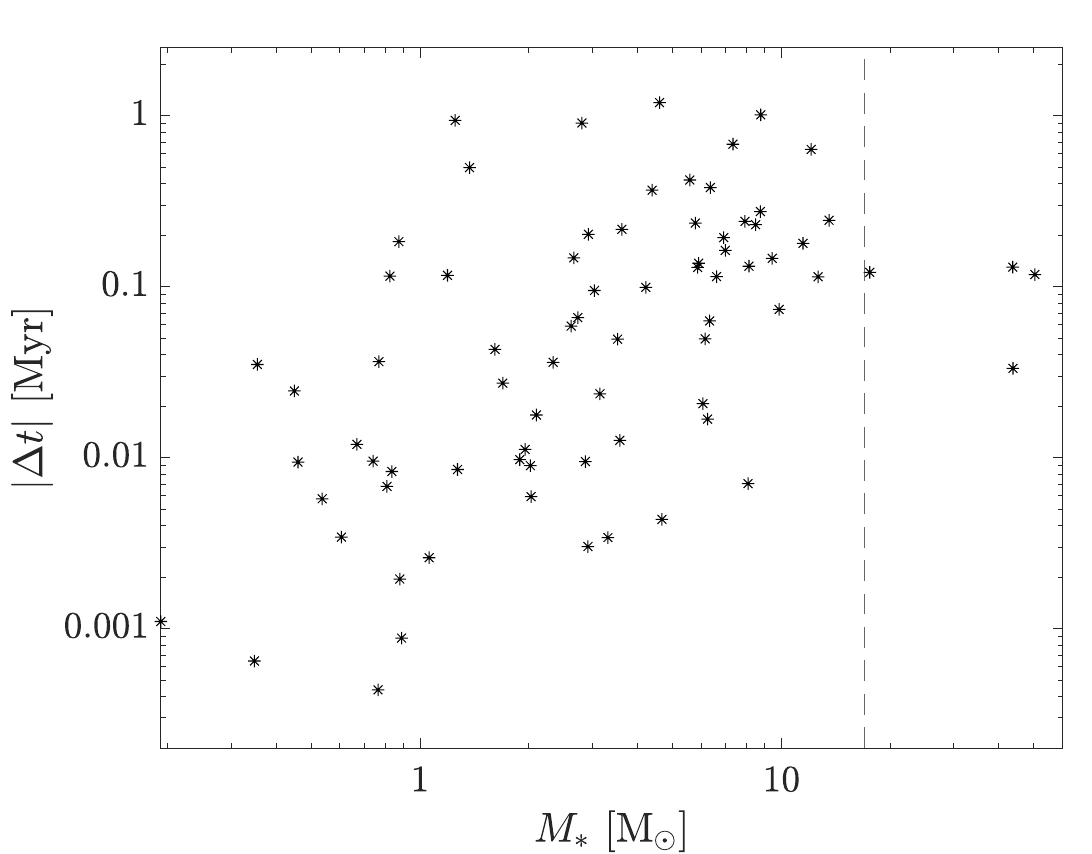}
    \includegraphics[trim={0.5cm 7.5cm 2cm 7.75cm}, clip, width=\hsize]{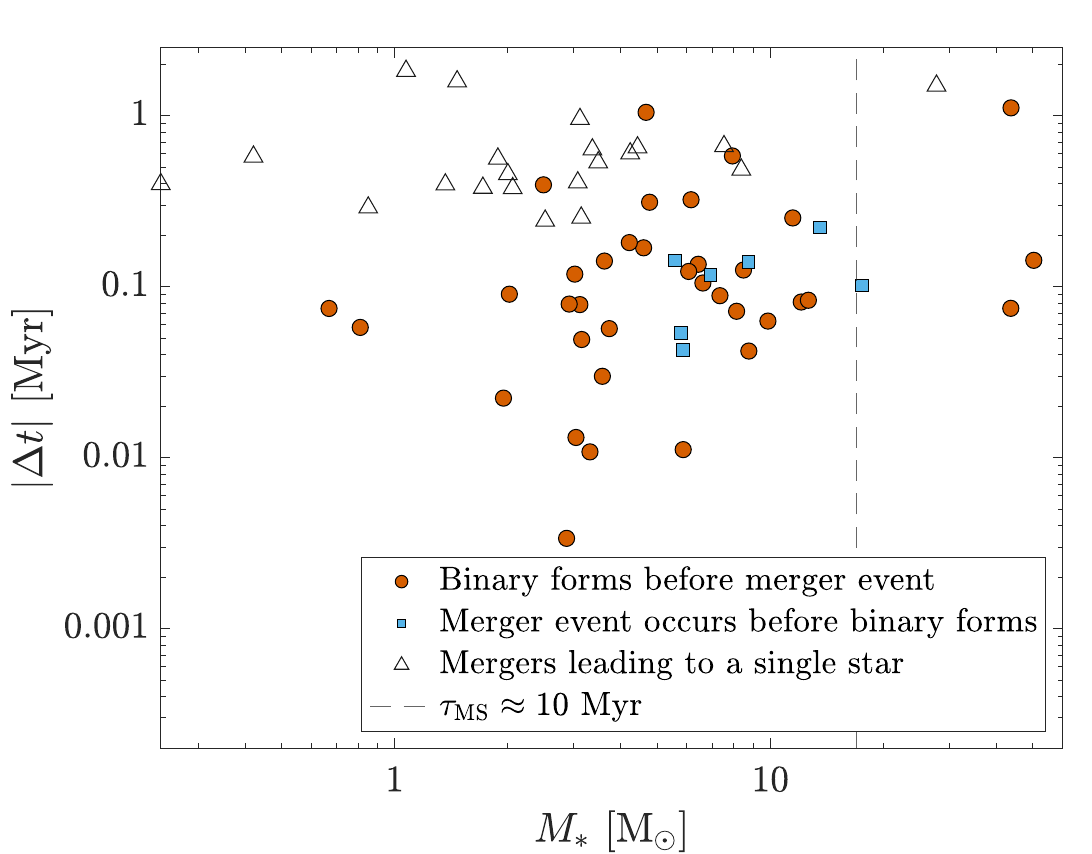}
    \caption{
    Delay time ($|\Delta t|$) between star formation (top panel) and binary assembly (bottom panel) as a function of the most massive star in the system.
    Top panel: black asterisks indicate the maximum time difference between the formation of the two (or three) stellar components that eventually comprise a binary (or a triple) system.
    Bottom panel: red circles represent systems in which the binary assembles before any merger event occurs, blue squares represent systems in which a merger occurs before the binary assembles, and black plus signs represent merger events that lead to single stars.    
    In both panels, a vertical dashed line indicates the approximate mass at which the main-sequence lifetime is 10 Myr.
    }
    \label{fig:delta_t_formation_and_merger}
\end{figure}

Figure~\ref{fig:delta_t_formation_and_merger} highlights two key timescales in our simulation: star formation within multiple systems is largely coeval (typically within 10--100 kyr), whereas mergers can occur much later, up to 1 Myr after formation. This late-time merging can bias age estimates of young massive stars if merger rejuvenation is not accounted for.

In the top panel, we display the maximum delay time in binary and triple systems. Most component stars of binaries or triples form within 10–100 kyr. Although some intermediate-mass stars exhibit formation delays of up to 1 Myr, their main-sequence lifetimes are approximately 10–100 Myr. For the most massive stars ($\approx 40-50 M_\odot$), the formation delay reaches about 0.1 Myr, which is only a few percent of their main-sequence duration. Overall, the simulation suggests that stars largely form nearly simultaneously.

In the bottom panel, we show the time elapsed between a system’s formation and its final merger event (red circles), cases in which the merger precedes binary formation (blue squares), and binary systems that eventually coalesce into a single star (black triangles). 
For some of the most massive stars, the delay between formation and merger can reach $\sim 1$ Myr, that is more than 10\% of their main-sequence lifetimes. Thus, even when stars form coevally, mergers may occur significantly later.
In our simulation, five stars have main-sequence lifetimes shorter than $10$ Myr, and two of these undergo mergers roughly $1$ Myr after their formation. This implies that $\sim 40\%$ of massive stars experience late-time mergers, highlighting that age estimates of young massive stars may be systematically biased if recent merger events are not accounted for.

\begin{figure*}
    \centering
    \includegraphics[trim={0cm 0cm 0cm 1.25cm}, clip, width=\hsize]{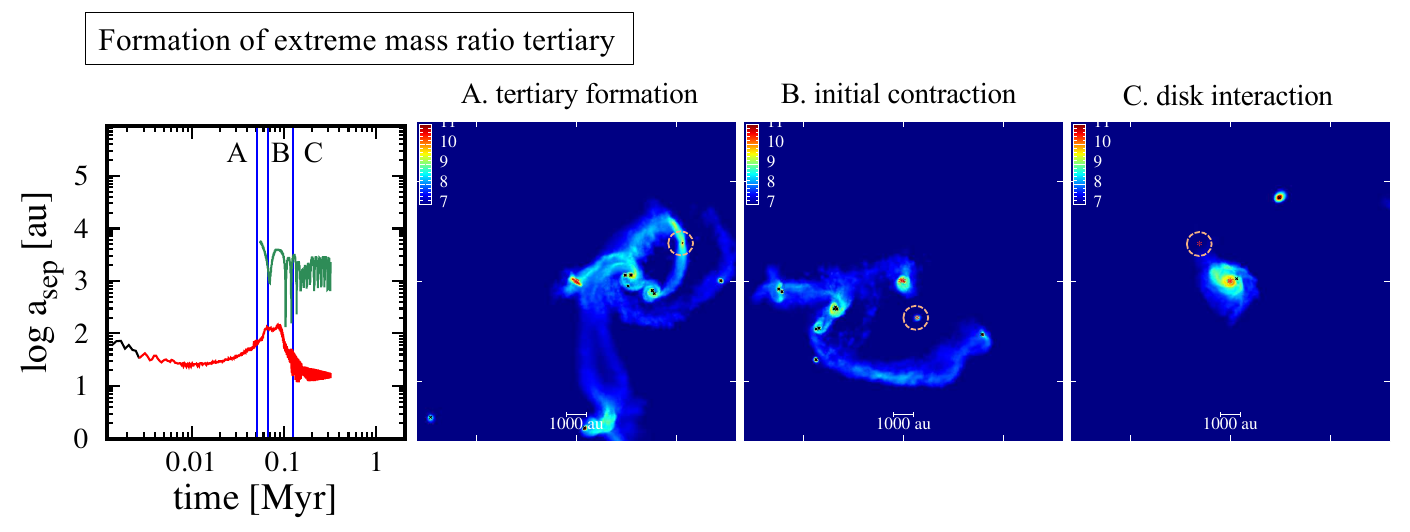}
    \caption{
    Example of a system in which a tertiary star forms with an extreme mass ratio of $M_{*,3}/(M_{*,1}+M_{*,2})\approx0.01$ (system ID 15).    
    This system has masses similar to those of the observed extreme mass-ratio binary reported by \citet{2024AJ....168..209P}.
    The dashed circles indicate the region where the tertiary will form (panel A) and position of tertiary (panels B and C).
    The red asterisks indicate the stars that eventually form a three-body system.
    The low mass tertiary is highlighted with a dashed circle.
    }
    \label{fig:extreme_mass_ratio}
\end{figure*}

\begin{figure}
    \centering
    \includegraphics[trim={0cm 0.15cm 0cm 0cm}, clip, width=\hsize]{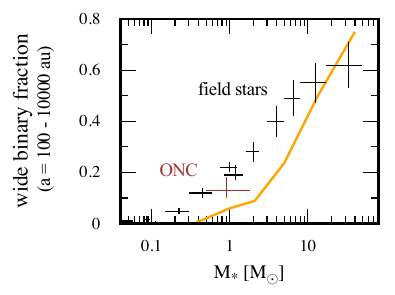}
    \caption{
    Wide binary fraction whose semi-major axis are in the range between $100~$ -- $10^4~$au. The line shows the simulated result.
    Black and red points indicate the observed values for the field binaries and binaries in Orion Nebula Cluster \citep{2018MNRAS.478.1825D, 2023ASPC..534..275O}, respectively.
    }
    \label{fig:wide_binary}
\end{figure}

\subsection{Extreme Mass Ratio}\label{subsec:disc:extreme_mass_ratios}
In our simulation, we identified 10 binaries with mass ratios $q<0.2$ (Figure \ref{fig:q_afinal}). 
Among these, three (or one) systems exhibit $q<0.1$ and have an intermediate-mass (or high-mass) primary. 
Overall, extreme mass-ratio ($q<0.1$) configurations appear more frequently in systems with massive primaries. 
Additionally, some massive triple systems include a low-mass tertiary. 
For instance, the most massive system in our simulation—a binary composed of $50.3 M_\odot$ and $48.5 M_\odot$ stars with $a_{\rm final} \approx 902$ au and $e = 0.60$—has a tertiary star of $0.4M_\odot$ located at $\approx 22,740$ au with $e = 0.69$ (Figure \ref{fig:merged_binary_fig} shows the assembly of this system).

Observationally, systems with extreme mass ratios are challenging to detect because the brighter component's luminosity overshadows that of its companion.
However, recent developments have led to their identification.

\cite{2024AJ....168..209P} reports two systems in Sco OB1 with low-mass companions:
i) O9.7IV star ($\sim 17 M_\odot$), possibly in a several day period binary, with a $\sim 0.13 M_\odot$ companion at $\sim$1110 au, and ii) B0.5IV star ($\sim 16 M_\odot$) with a $\sim 0.3 M_\odot$ companion at $\sim$790 au.
While our simulation does not produce an exact analogue of a single OB star with a wide, very low-mass companion, we do find a qualitatively similar configuration in the form of an extreme mass-ratio triple.
We identify such a triple system in which the inner binary has a combined mass of $\approx 12.5\,M_\odot$ and is orbited by a very low-mass tertiary of $0.1\,M_\odot$ (system ID 15). The inner binary is moderately wide, with a final separation of $a_{\rm final}\approx 14$ au and eccentricity $e\approx 0.11$, whereas the tertiary orbits at $a_{\rm final}\approx 2300$ au with $e\approx 0.62$. 
Figure~\ref{fig:extreme_mass_ratio} illustrates the formation pathway of this extreme mass-ratio triple system.

All three stars originate within the same progenitor core. By the time the tertiary forms, the other two stars have already assembled into a relatively tight binary with a separation of $\lesssim 100$ au. As the binary and tertiary migrate within the collapsing core, their orbits approach to pericentric distances of several hundred au.  During these close passages, the tertiary interacts with the circumbinary material surrounding the inner pair and accretes mass from the disc.  
However, the circumbinary disc is rapidly dispersed by strong UV radiation from the massive inner binary, limiting the mass accretion onto the tertiary. As a result, the tertiary remains a low-mass star, leaving behind an extreme mass-ratio hierarchical system.

\citet{2025A&A...703A.239N} recently reported a sample of newly identified systems consisting of a B-type primary ($7 \lesssim M_*/M_{\odot} \lesssim 13$) in a short-period ($\sim 1$ d) orbit with a low-mass companion of order $\sim 1\,M_{\odot}$.
In our simulation, one of the binaries that subsequently undergoes a merger exhibits an extremely small mass ratio of $q=0.06$.
If this system survived without merging and tightened to sub-au separations, it would have represented a plausible progenitor of the observed extreme–mass-ratio systems.
Resolving such binaries, however, requires substantially higher numerical resolution.
Future simulations capable of following separations below $\sim 0.1$ au—and including the relevant physics such as tidal dissipation and mass transfer—will be essential for assessing the survivability, orbital evolution, and formation pathways of massive binaries with extreme mass ratios.

\subsection{Wide Binaries}
At the end of our simulation, we find 16 (or 12) binaries with separations larger than $10^3$ (or $10^4$) au (Figure \ref{fig:ainit_afinal}).
Both low-mass ($M_*\leq2 M_\odot$) and intermediate- to high-mass ($M_*>2 M_\odot$) systems contribute to the formation of these binaries.
Very wide binaries ($a_{\rm final}\gtrsim 10^5$ au) are slightly more common among low-mass stars. However, these very wide binaries are loosely bound and vulnerable to disruption (e.g., through prolonged simulation times, fly-bys, or the Galactic tide). Additionally, there is a tentative correlation between final separation and eccentricity (Figure \ref{fig:ecc_afinal}), which may indicate a super-thermal distribution at birth \citep[e.g.,][]{2019ApJ...872..165G,2024MNRAS.532.2425H,2024MNRAS.532.2374M}.

Figure~\ref{fig:wide_binary} shows the wide-binary fraction as a function of primary stellar mass. 
Our simulated clusters reproduce the observed trend that the wide-companion fraction increases steadily with primary mass, a behaviour well established for B- and O-type stars in both young clusters and the field \citep[e.g.][]{2023ASPC..534..275O}. 
For primaries with $M_* \gtrsim 10\,M_\odot$, the wide-binary fraction in our simulation reaches $50$–$60\%$, consistent with observational estimates for massive stars. 
At lower primary masses ($M_* \lesssim 10\,M_\odot$), the simulated values are slightly below the observational estimates, though the overall mass dependence and normalisation remain in good agreement with the data.

The mass-ratio distribution of these wide companions also agrees well with observations. 
We find a median mass ratio of $q \simeq 0.28$, close to the characteristic value $q \simeq 0.2$ inferred for wide companions to mid-B and O-type stars \citep{2018A&A...620A.116G, 2023ASPC..534..275O}. 
We note that these wide binaries mostly originate from core and disc fragmentation (Table~\ref{tab:formation_mode}). 
Such low mass ratios are broadly consistent with expectations from core or filament fragmentation, in which the tertiary forms from its own mass reservoir rather than coevolving within a shared circumbinary disc. 
This naturally yields mass ratios similar to those drawn randomly from a Salpeter-like initial mass function.

The agreement between our simulated wide-binary population and the observed distributions strengthens the robustness of our analysis. 
Wide companions are primarily governed by large-scale gravitational fragmentation and remain largely insensitive to the details of the numerical methodology—such as spatial resolution, sink accretion prescriptions, or magnetic-field treatment—which predominantly affect the formation and evolution of tight systems (Section~\ref{sec:caveats}).

\subsection{Massive Triples}
We identify 24 hierarchical triple systems among the 76 multiple systems formed in our simulation. These triples are associated with massive primary stars: only one system has a primary mass below $2\,M_\odot$ (see also Fig.~\ref{fig:multiplicity}). We interpret this trend as a consequence of the formation environment: more massive stars tend to originate from more massive parental cores, which fragment into a larger number of stars and thus provide more opportunities to build higher-order multiples. This interpretation is consistent with observational evidence that the tertiary fraction increases with the mass of the primary star.

\begin{figure}
    \centering
    \includegraphics[trim={0.5cm 7.5cm 1.0cm 7.75cm}, clip, width=\hsize]{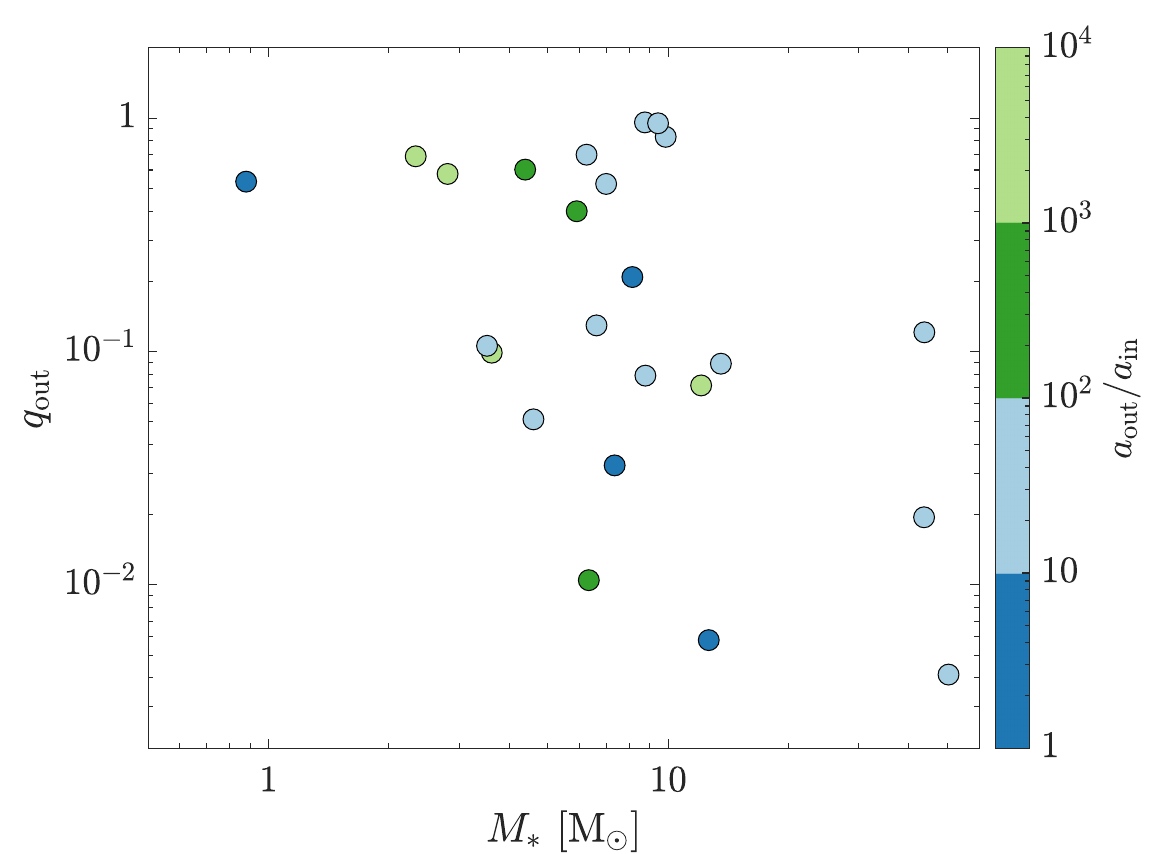}
    \caption{
    All bound triple-star systems at the end of our simulation.
    Each triple is characterised by the mass of the most massive star in the system ($M_{*}$), the outer mass ratio ($q_{\rm out}=M_{*,3}/(M_{*,1}+M_{*,2}))$, and coloured by the outer-to-inner separation ratio ($a_{\rm out}/a_{\rm in}$).
    }
    \label{fig:massive_triples}
\end{figure}

In our sample, the tertiary companion is frequently the most massive component: in 10 out of 24 triples, the tertiary star is the most massive member. However, in none of these systems does the tertiary mass exceed the combined mass of the inner binary (i.e. $M_3 < M_1+M_2$), a configuration that has been reported in massive stellar systems \citep{2022MNRAS.511.4710E}. Figure~\ref{fig:massive_triples} shows all triple systems at the end of the simulation, demonstrating that the tertiary-to-inner-binary mass ratio $q_{\rm out}\equiv M_3/(M_1+M_2)$ never exceeds unity in our sample, while the distribution clusters near unity, qualitatively resembling the rise toward $q\simeq 1$ for binary mass-ratio distributions (Fig.~\ref{fig:q_afinal}). Establishing this behaviour quantitatively, however, will require a larger sample of triple systems.

Figure~\ref{fig:massive_triples} indicates that most triples in the final snapshot are dynamically stable, with $a_{\rm out}/a_{\rm in}>10$. This is expected, since the majority of these systems formed more than $0.1,{\rm Myr}$ earlier, allowing enough time for dynamically unstable configurations to be disrupted (Fig.~\ref{fig:separation_evolution}). 
Only four systems may be susceptible to dynamical instability because the outer orbit is insufficiently hierarchical, with $a_{\rm out}/a_{\rm in}\lesssim5$ \citep{2001MNRAS.321..398M}. Two of these systems have extremely wide outer separations ($a_{\rm out}\sim10^5\,{\rm au}$), so any instability may not yet have developed owing to their long outer dynamical times. The only sub-solar-mass triple in our sample, however, lies in the dynamically unstable regime with the final separation of $a_{\rm in} \sim 3000~$au, implying that the already small population of low-mass triples may be further depleted as the system evolves.

\subsection{Caveats} \label{sec:caveats}
\subsubsection{Magnetic fields}
Our simulation does not include magnetic fields, which could have a non-negligible impact on the formation and evolution of massive tight binaries. Magnetic fields are known to significantly alter the structure and dynamics of circumstellar discs. For example, a coherent magnetic field can remove angular momentum from the disc gas via magnetic braking, potentially reducing the disc size if the gas remains well-coupled to the magnetic field \citep[e.g.,][]{2008ApJ...681.1356M, 2011PASJ...63..555M, 2013ApJ...766...97M}. Turbulent magnetic fields may also stabilize disc structures by providing additional support through magnetic pressure, thereby suppressing fragmentation \citep[e.g.,][]{2020MNRAS.497..336S}. Moreover, magnetic pressure can influence the location of the inner edge of circumstellar discs and stabilize dense cores and filaments, reducing the overall star formation efficiency.

Despite the growing recognition of the importance of magnetic fields, global star formation simulations that consistently incorporate both magnetic fields and non-ideal magnetohydrodynamic (MHD) effects, such as Ohmic dissipation, ambipolar diffusion, and the Hall effect, remain limited. For instance, \citet{2021MNRAS.502.3646G} include magnetic fields in their simulations, but do not account for non-ideal MHD effects. This likely leads to unrealistically strong magnetic braking, which may suppress the formation of circumstellar discs.
In future work, incorporating non-ideal MHD physics will be essential for accurately capturing the formation and evolution of circumstellar discs, especially around massive stars. Recent studies have demonstrated that non-ideal effects can weaken magnetic braking and allow disc formation even in strongly magnetised environments \citep[e.g.,][]{2015ApJ...801..117T, 2015MNRAS.452..278T, 2016MNRAS.457.1037W, 2018A&A...615A...5V, 2025MNRAS.543.3321M}. 
Including these processes in high-resolution global simulations could provide a more comprehensive understanding of how magnetic fields influence the formation of binary populations.

\subsubsection{Spatial Resolution}\label{subsubsec:disc:resolution}
While our simulation achieves the highest spatial resolution among existing studies \citep[e.g., Figure 6 of][]{2024ARA&A..62...63H}, it still suffers from resolution effects, especially in modelling the formation and evolution of massive tight binaries with separations of $a \lesssim 10~\mathrm{au}$ \citep[e.g.,][]{2025ApJ...989..134R}. Our gravitational softening length is $0.2~\mathrm{au}$, and we resolve gas densities up to $\sim 2 \times 10^{15}~\mathrm{cm^{-3}}$. These values are comparable to or higher than those used in previous studies; for example, \citet{2019MNRAS.484.2341B} adopted a softening length of $0.5~\mathrm{au}$, while \citet{2023MNRAS.518.4693G} used $7.56~\mathrm{au}$.
Spatial resolution can significantly impact the evolution of binary systems around the sink radius, where star–disc interactions efficiently drive orbital migration. However, sink accretion artificially reduces the gas density near the sink, thereby decelerating inward migration. 
Furthermore, ionizing radiation, particularly from massive stars, photo-evaporates circumstellar discs from their inner edges. In high-resolution runs, denser regions near the disc edge may be better resolved, leading to stronger shielding against ionizing photons and thus hindering photo-evaporation. 
These resolution-dependent effects likely lead to an underestimation of the migration rate, particularly for massive binaries.

We include mergers between sink particles in our simulation. When the binary separation becomes smaller than the sink radius, we assume the two stars merge—even though their true stellar radii are much smaller. This approach may introduce uncertainties, especially for massive stellar systems where merger events are more common. For instance, i) some events that we classify as mergers might actually result in the formation of tight binaries, ii) the multiplicity fraction in our simulation may be underestimated, and iii) the resulting mass spectrum might be artificially top heavy.
To assess the impact of resolution on our mass function estimation, we present both the original and “merger-corrected” mass spectra (Fig. \ref{fig:IMF} and Section~\ref{subsubsec:correct_mergers}); overall, both remain consistent with the Salpeter IMF. Notably, previous studies \citep[e.g.,][]{2019MNRAS.484.2341B,2023MNRAS.518.4693G} did not allow for stellar mergers in their simulations, so their statistics on tight massive binaries cannot be directly compared to ours. Future simulations with smaller sink radii or more refined merger criteria will be necessary to better assess the formation and survival of tight massive binaries.

\section{Summary and Conclusions} \label{sec:conclusion}
We have studied the origin of the binary system, analysing the results of the radiation hydrodynamic simulation at Solar metallicity first presented in \citet{2024MNRAS.530.2453C}. Our main findings are summarized as follows.

\begin{description}
    \item \textbf{Binary formation and early migration.} 
    Massive binaries in the simulation assemble through four primary channels: filament fragmentation, core fragmentation, disc fragmentation, and dynamical capture. Dynamical captures lead to the widest systems, with separations exceeding $10^{4}$ au, while tight binaries ($a < 100~$au) originate almost equally from filament, core, and disc fragmentation.
    Each formation channel sets a characteristic initial separation: disc fragmentation typically yields binaries at a few hundred au, whereas filament and core fragmentation produce systems separated by a few thousand au. However, regardless of the channel, newly formed binaries experience rapid early migration, with their separations shrinking by more than an order of magnitude within the first $0.1$–$0.2$ Myr. By the end of the simulation, the separation distribution spans smoothly across $1$–$10^{4}$ au, indicating that the memory of the initial conditions has largely been erased.
    \item \textbf{Formation of tight binaries.}
    Tight binaries with final separations below 10 au form through three distinct evolutionary phases: an initial contraction phase, a disc–star interaction phase, and a circumbinary-disc phase.
    In the initial contraction phase, stars born within the same core move closer together as the collapsing progenitor core’s self-gravity and repeated multi-body scatterings draw them inward.
    When the pericentre distance decreases to a few hundred au, the system enters the disc--star interaction phase.  
    This interaction excites prominent spiral arms in the discs, efficiently extracting orbital angular momentum in a manner analogous to Type-I migration.
    As the binary contracts to separations of several tens of au, both stars become embedded in a common disc. Sustained torques from this circumbinary material gradually harden the orbit. 
    Notably, all binaries in our simulations that reach final separations below 10 au undergo this phase.   
    \item \textbf{Multiplicity.}
    The binary and tertiary fractions as functions of primary stellar mass in our simulation reproduce the observed trend: the binary fraction increases from $\sim 0.1$ at $M_* \sim 0.1\,M_\odot$ to nearly unity for $M_* \gtrsim 10\,M_\odot$. Although the qualitative behaviour matches observations, the simulation underproduces low-mass binaries. A comparison with a similar run without radiative feedback shows much better quantitative agreement, suggesting that more detailed radiative-transfer treatment within circumstellar discs may be necessary to accurately reproduce the observed low-mass binary population.
    \item \textbf{Mergers.}
    Stellar mergers frequently occur during cluster assembly. On average, about 15\% of stars experience at least one merger, and massive stars often undergo multiple mergers since they predominantly form in high-density regions. The typical merger timescale ranges from 0.1 to 1 Myr. Notably, we have identified two systems that merge more than 1 Myr after formation, even though their stellar lifetimes are less than 10 Myr. These delayed mergers could introduce significant uncertainties in stellar age estimates.
    In contrast, the delay between star formation events in a binary system is relatively small, allowing the stars to be treated as coeval and minimising any significant uncertainties in age estimates.
    
    \item \textbf{Orbital orientation.}
    The orbital axes of binary systems are randomly oriented and show no correlation with the initial rotational axis of the parent cloud. The mutual inclinations between inner and outer orbits in hierarchical systems are likewise consistent with random orientations. This isotropy suggests that the formation of massive binaries is driven by turbulent, stochastic processes with cloud turbulence and multi-body interactions playing key roles in shaping their final orbital configurations.
    \item \textbf{Extreme mass ratio binaries and wide binaries}
    We also find the extreme mass ratio binaries with the symmetric mass ratio below $0.1$. They mostly have the separation larger than $100$--$1000~$au, where disc–star interactions occur only for a limited time because the discs are rapidly dispersed by strong feedback. 
    We have found one merged system with the extremely mass ratio of $q=0.06$. If these stars survive as the tight binaries within the sink radius of $\sim 1~$au, this provides analogous sample to recently observed tight and extreme mass ratio binary \citep{2025A&A...703A.239N}.
\end{description}

\section*{Acknowledgements}
The authors thank Selma de Mink, Dan D'Orazio, Harim Jin, Jakub Klencki, Norbert Langer, Ilya Mandel, Max Moe, Tinne Pauwels, Mark Krumholz, Masato Kobayashi, Abinaya Rajamuthukumar, Chris Tiede, Long Wang, and Hans Zinnecker for useful discussions and helpful suggestions. 
We use the SPH visualization tool SPLASH \citep{2007PASA...24..159P} in Figs.~\ref{fig:overall}, \ref{fig:circum_binary_fig}, \ref{fig:merged_binary_fig}, and \ref{fig:extreme_mass_ratio}.

\section*{Data Availability} 
The data underlying this article will be shared on reasonable request to the corresponding author.

\bibliographystyle{mnras}
\bibliography{references}



\bsp	
\label{lastpage}
\end{document}